\documentclass[useAMS,usenatbib]{mn2e}
\usepackage{natbib}
\usepackage{aas_macros}
\usepackage{graphicx}
\usepackage{multirow}
\usepackage{tabularx}
\usepackage{gensymb}
\usepackage{amsmath}

\begin{document}
\title[sub-L* galaxy evolution]{The 5 Gyr evolution of sub-M* galaxies}
\author[K. Disseau et al.]
  {
  K.~Disseau$^{1}$, M.~Puech$^1$, F.~Hammer$^1$, H.~Flores$^1$, Y.~Yang$^1$, and M.~Rodrigues$^1$ \\
  	$^1$ GEPI, Observatoire de Paris, PSL Research University, CNRS, Univ. Paris Diderot, Sorbonne Paris Cit\'e,\\ Place Jules Janssen, 92195 Meudon, France\\
   }
  
\date{Released 2002 Xxxxx XX}

\pagerange{\pageref{firstpage}--\pageref{lastpage}} \pubyear{2015}

\maketitle

\label{firstpage}

\begin{abstract}
  We gathered two complete samples of $M_{AB}(r)<-18$ ($M_{star} >
  10^{9} M_{\odot}$) galaxies, which are representative of the
  present-day galaxies and their counterparts at 5 Gyr ago. We
  analysed their 2D luminosity profiles and carefully decomposed them
  into bulges, bars and discs. This was done in a very consistent way
  at the two epochs, by using the same image quality and same (red)
  filters at rest. We classified them into elliptical, lenticular,
  spiral, and peculiar galaxies on the basis of a morphological
  decision tree.

  We found that at $z=0$, sub-M* ($10^{9} M_{\odot} <
  M_{star} < 1.5 \times 10^{10} M_{\odot}$) galaxies follow a similar
  Hubble Sequence compared to their massive counterparts, though with
  a considerable larger number of (1) peculiar galaxies and (2) low
  surface brightness galaxies. These trends persist in the $z\sim0.5$
  sample, suggesting that sub-M* galaxies have not reached yet a
  virialised state, conversely to their more massive counterparts. The
  fraction of peculiar galaxies is always high, consistent with a
  hierarchical scenario in which minor mergers could have played a
  more important role for sub-M* galaxies than for more massive galaxies.

  Interestingly, we  also discovered that more than 10\% of the
  sub-M* galaxies at z=0.5 are low surface brightness galaxies with
  clumpy and perturbed features that suggest merging. Even more
  enigmatic is the fact that their disc scale-lengths are comparable to
  that of M31, while their stellar masses are similar to that of the
  LMC.
\end{abstract}


\section{Introduction}

The deepest HST images revealed a plethora of intrinsically faint
galaxies at z $\sim$ 1 (\citealt{Ryan07}). Because the
  $\Lambda$-CDM model predicts that galaxies are assembled over cosmic
  times through the hierarchical merging of progressively more massive
  sub-units (e.g., \citealt{white78}), possibly combined with massive
  accretion of cold gas (e.g., \citealt{keres09}, although see
  \citealt{nelson13}), it is then expected that the fraction of dwarf
  galaxies amongst the whole galaxy population increases as a function
  of redshift $z$. This should be reflected by a steepening of the
  faint-end slope of the galaxy LF. However, observational results are
  quite puzzling. On the one hand, first results based on rest-frame
  UV data indeed detected such an increase (e.g., \citealt{Ryan07},
  \citealt{reddy09}). But on the other hand, more recent results at
  rest-frame optical wavelengths revealed that the faint-end slope
  actually appears to remain constant out to $z \sim 4$ (e.g.,
  \citealt{marchesini12}), hence quite in contrast with expectations
  from the $\Lambda$-CDM model \citep{khochfar07}.

Distant sub-M* galaxies (i.e., galaxies with masses smaller
  than the knee of the LF) have faint apparent magnitudes ($z_{AB}$ $\sim$
  24 at $z=0.5$) that have limited detailed morphological studies and
  3D spectroscopy with current facilities until recently. Deep recent
  HST surveys and 3D observations have now started to investigate
  their morphological properties (e.g.,
  \citealt{vanderwel14,karta15,whitaker15}) and kinematic properties
  (e.g., \citealt{kassin12,simons15,contini16}).
Besides spirals and ellipticals, other types of galaxies may
increasingly appear in the sub-M* range, as mass decreases from $\sim
10^{10} M_{\odot}$ to $\sim 10^{9} M_{\odot}$. These include low
surface brightness galaxies (see \citealt{Zhong12} and references
therein), tidal dwarf galaxies (TDGs, see Fig. 2 of
\citealt{Kaviraj11}), and objects more and more similar to dwarfs
(i.e., elliptical: dE, spiral: dSp, and irregular: dIrr, see, e.g.,
\citealt{Kormendy09}). Spirals in this mass range form a sequence in
the surface brightness vs. absolute magnitude plane together with Irrs
and dSphs (see \citealt{Kormendy09}, and references therein), which
strongly differs from the sequence formed by Ellipticals and dEs.

The goal of this paper is to investigate the detailed morphology
  of sub-M* galaxies and investigate possible formation
  mechanisms in comparison with super-M* galaxies. For this, we
  followed the careful methodology introduced by \cite{Delgado10}
for $M_{stellar}>1.5\times10^{10}$ $M_{\odot}$ galaxies. We present a
complete morphological inventory of these galaxies at two epochs
(i.e., 0 and 5 Gyr ago), as described in Sect. 2. Methodology and
analysis are detailed in Sect. 3. Sect. 4 investigates more
specifically low surface brightness galaxies, while a complete
morphological classification is presented in Sect. 5. We discuss in
Sect. 6 the accuracy of the results and their main limitations, and
conclusions are drawn in Sect. 7. Throughout this paper we adopt
cosmological parameters with $H_0=70$ km.s$^{-1}$.Mpc$^{-1}$ and
$\Omega_{\Lambda}$=0.7. Magnitudes are given in the $AB$ system. 

\section{Representative samples of nearby and distant galaxies}

\subsection{Methodology \& sample sizes}

The morphological classification was done following the
  classification tree introduced by \cite{Delgado10}. This requires
  two main steps: (1) the analysis of the light distribution and its
  decomposition into several components (bulge, disc, and bar when
  necessary), and (2) the analysis of the galaxy color map. These
  steps are detailed in Sect 3. The global uncertainty associated to
  any classification is a combination of Poisson random fluctuations
  due to the finite size $N$ of the sample and possible systematics
  due to the intrinsic relative (in)accuracy of the classification
  process (i.e., the fraction of objects that are not properly
  classified as a result of approximations in the methodology used
  during the classification, see Sect. 3). While the random
  uncertainty varies as $\sqrt{N}/N$, i.e., it decreases with larger
  samples, the systematics are independent of the sample size. Using
  the \cite{Delgado10} classification tree, we estimated the latter to
  be $\sim$ 10\% from a comparison between morphology and
  spatially-resolved kinematics (see \citealt{Neichel08}). This
  implies that the total uncertainty (random and systematics) will
  remain limited by systematics regardless of the sample size once $N$
  reaches $\sim$100 galaxies. Note that automatic classification
  methods lead to even larger systematics (e.g., \citealt{Neichel08},
  \citealt{povic15}). The sample sizes were therefore fixed to $N=150$
  objects.

\subsection{Sample selection}
\subsubsection{Local galaxy sample}

\begin{figure}
\begin{center} \includegraphics[width=80mm]{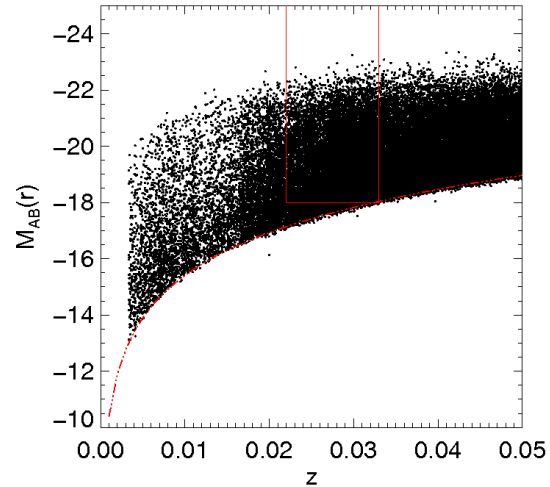} \end{center}
\caption{$M_{AB}(r)-z$ plane for the SDSS galaxies. The region
  delimited by the red solid lines delineates the local volume-limited
  sample with $M_{AB}(r) \leq -18$ and 0.022$\leq z \leq$0.033.}
\label{Mrzplane}
\end{figure}

Galaxies were selected from the Sloan Digital Sky Survey Release Seven
(SDSS DR7, \citealt{Abazajian09}). We used the low-$z$ catalogue of
the New York University Value Added Galaxy Catalog (NYU-VAGC,
~\citealt{Blanton05}), which includes absolute magnitudes,
$k$-corrections, and extinctions in the $u$ (3551\AA), $g$ (4686\AA),
$r$ (6165\AA), $i$ (7481\AA), and $z$ (8931\AA) bands.
Figure~\ref{Mrzplane} shows the relationship between redshift and
absolute magnitude in $r$ band, from which we extracted an unbiased
volume-limited sample: given a cut in luminosity ($M_{AB}(r)\leq -18$)
a redshift cut ($z\leq0.033$) is necessary to avoid any Malmquist
bias. We also defined a lower redshift cut ($z\geq0.022$) to ensure
that the peculiar motions of the closest galaxies could not affect
redshift and distance determinations. At this stage the parent sample
contained 12,681 galaxies. A representative subsample of 150 galaxies
was then selected at random. One galaxy was found to fall on the CCD
edge, which was removed from the final sample. Figure~\ref{lumf}
compares the distribution of the $r$-band absolute magnitudes for the
final sample of 149 galaxies with the $r$-band luminosity function of
local galaxies~\citep{Blanton03}. A Kolmogorov-Smirnov test indicates
a probability of 99\% that both the sample and the local luminosity
function are drawn from the same parent distribution. This sample is
thus representative of the $M_{AB}(r)\leq-18$ galaxy population in the
local Universe. We checked that other properties such as the
   $g-r$ color and the Petrosian half-light surface brightness
  (extracted from the NYU-VAGC catalogue) in the final sample and parent
  samples also follow the same distributions. Luminosities in $r$ band
were converted into stellar masses $M_{stellar}$ using the statistical
empirical relation between $M_{stellar}/L_r$ and the $u-g$ color from
\citet{Bell03}.

\subsubsection{Distant galaxy sample}

\begin{figure*}
\includegraphics[width=80mm]{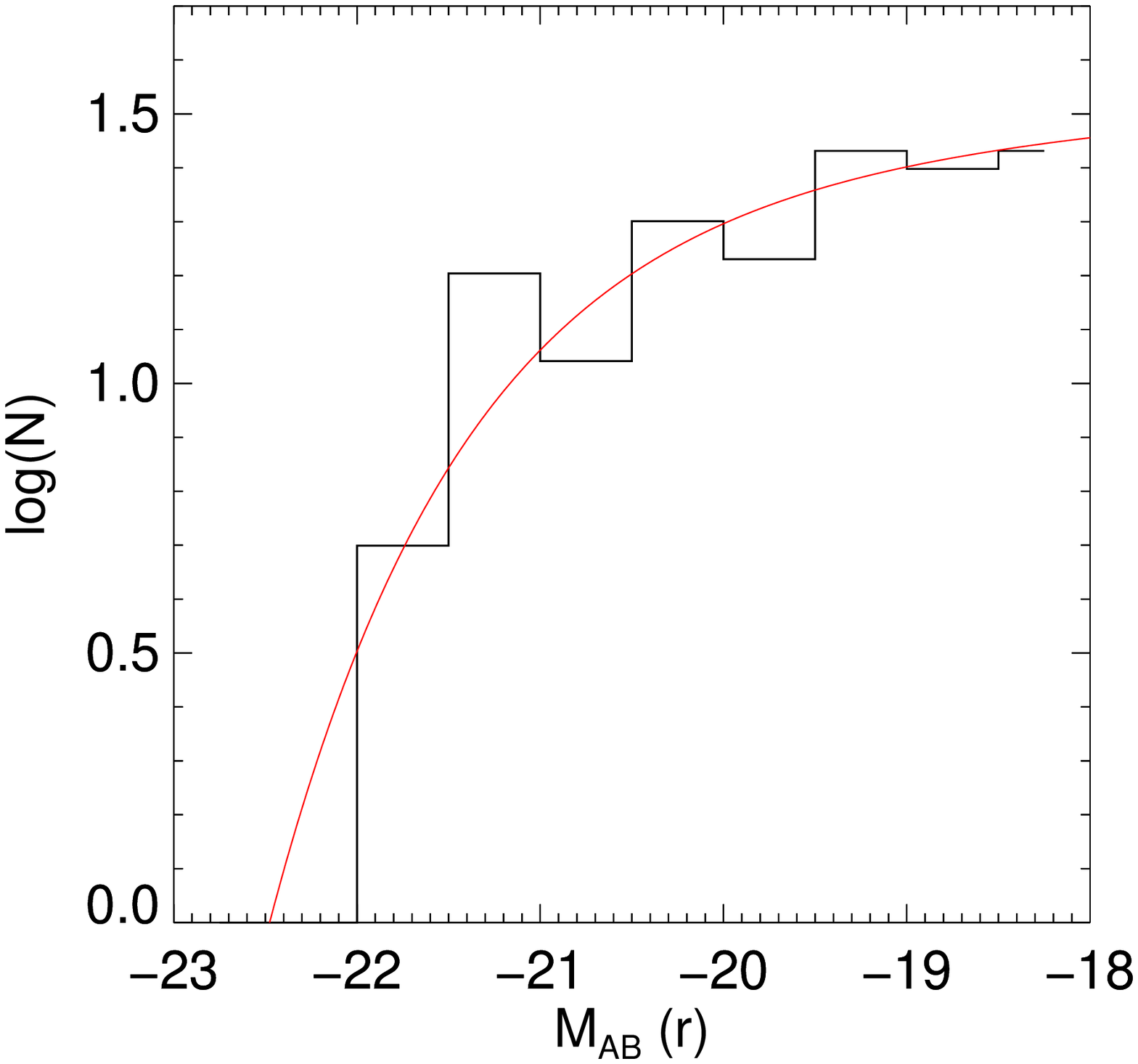}
\includegraphics[width=80mm]{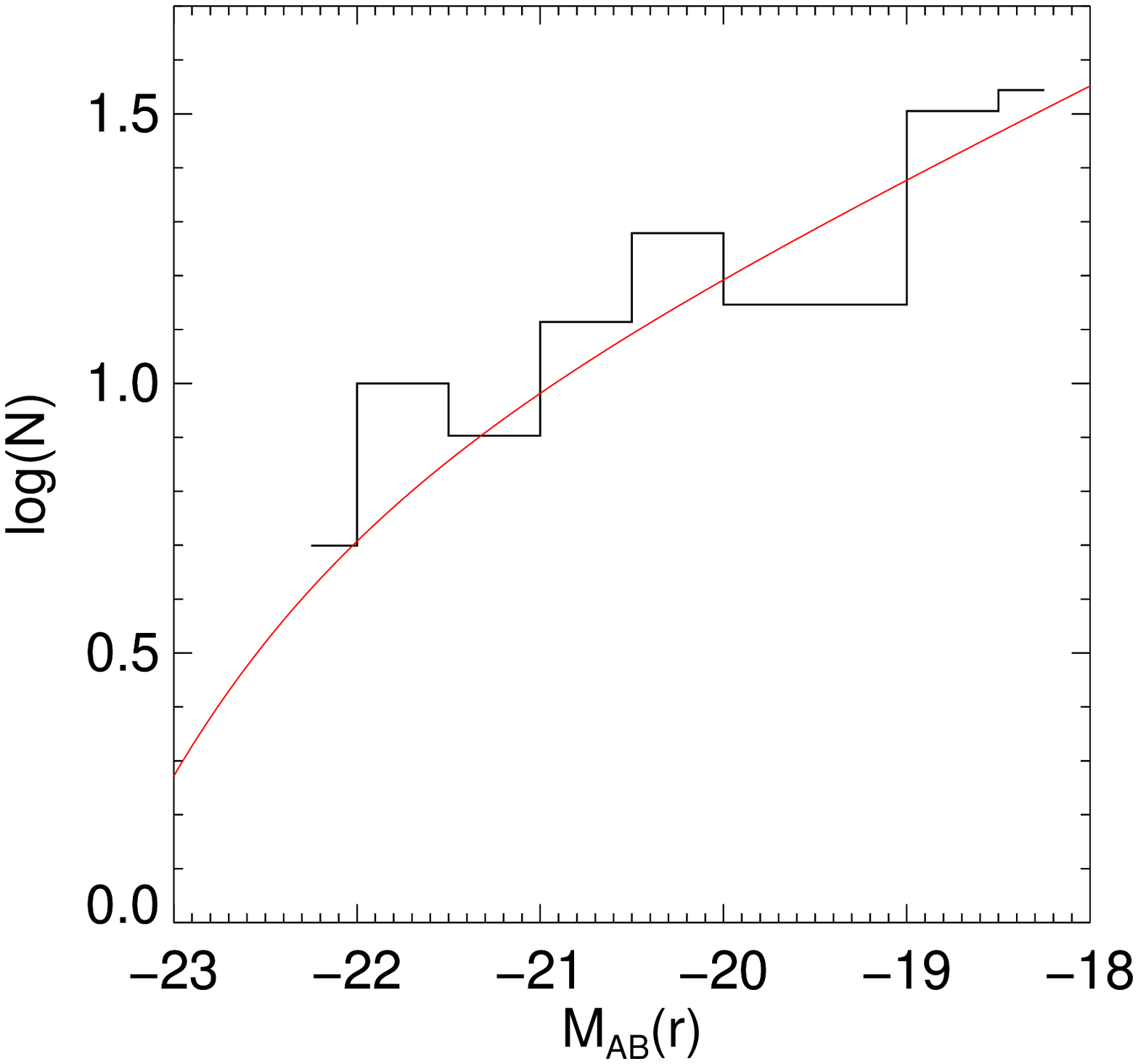}
\caption{\textit{Left :} $M_{AB}(r)$ histogram of the 150 galaxies in
  the local sample. The observed luminosity function from
  \citet{Blanton03} is over-plotted in red. \textit{Right :}
  $M_{AB}(r)$ histogram of the 150 galaxies in the distant sample with
  the observed red band luminosity function from \citet{Ilbert05}
  shown as a red line.}
\label{lumf}
\end{figure*}

The distant sample was selected from the CANDELS/GOODS-South survey
(\citealt{Koekemoer11}, \citealt{Grogin11}). Using SExtractor \citep{Bertin96} in the
HST/ACS $B$ (4312\AA), $V$ (5915\AA), $i$ (7697\AA), and $z$ (9103\AA)
v2.0 images and in the HST/WFC3 $J$ (12486\AA) and $H$ (15369\AA) v1.0
images, we built our own photometric catalogues, which were then
cross-correlated with the catalogue of photometric redshifts
from~\citet{Dahlen10}. Stellar masses and absolute magnitudes were
calculated using the prescription discussed in \citet{Bell03} (see
also \citealt{Hammer05} and references therein). We applied the same
luminosity criterion we used in the local sample, i.e.,
$M_{AB}(r)\leq-18$, and selected sources with photometric redshifts
between 0.4 and 0.6, which ensures that the observed light in $z$ band
corresponds to a rest-frame emission in $r$ band. The 278 sources thus
selected were visual inspected in all bands and contaminating stars
were removed\footnote{Saturated stars were not separated from galaxies
  during the photometric redshift estimation; because absorption bands
  present in K star spectra can easily be confused with the 4000\AA{}
  break of a galaxy spectrum, there remained in the catalogue a large
  number of ``low redshift'' objects that were actually stars.} from
the sample, leading to a final sample of 229 galaxies with
$M_{AB}(r)\leq-18$ and $0.4\leq z_{phot}\leq 0.6$. 150 galaxies were
then selected from this parent sample at random. The comparison
between the $M_{AB}(r)$ distribution of the final sample and the
$R$-band luminosity function of galaxies at redshifts 0.4 to 0.6 from
\citet{Ilbert05} is shown in Fig.~\ref{lumf}. A Kolmogorov-Smirnov
test gives a probability of 99.9\% that the sample and the luminosity
function are drawn from the same parent distribution, which evidences
the representativity of the distant sample.

\subsection{Comparison between the SDSS and HST images}

The aim is to classify the morphology of two galaxy samples that
  are representative of the same galaxy population at two different epochs.
  To avoid potential biases while comparing samples at different
  redshifts, it is important to make sure that the spatial resolution,
  depth, and rest-frame wavelengths sampled by all the images are
  similar.

  The rest-frame wavelengths at which morphology is studied must be
  representative of galaxy mass, i.e., images in a photometric band
  that samples the galaxy spectrum $>$ 4000 \AA~ at rest are required.
  The SDSS $r$ band (with a central wavelength $\lambda _c=$ 6166 \AA;
  see Sect. 2.2.1) from which the local sample was selected satisfies
  this requirement, and also presents the advantage to be redshifted
  into the HST $z$ band at $z\sim0.5$.

The 0.1 arcsec PSF FWHM of the GOODS/ACS images corresponds to 0.63
kpc at the median $z=0.53$ of the distant sample, while the 1.4 arcsec
PSF FWHM of the $r$-band SDSS images corresponds to 0.81 kpc at the
median $z=0.029$ of the local sample, which implies that all the
images sample similar spatial resolution. Moreover, the good agreement
between the rest-frame bands sampled by the HST/ACS and the SDSS data
(see Tab.~\ref{tab:lambda}) minimizes the $k$ correction between the
two data sets. Finally, the depths of the two surveys can be
  compared using the ratio between the two signal-to-noise ratios that
  would be obtained if one would be observing the same source at $z=0$
  from the SDSS and at $z=0.5$ from GOODS (see detailed calculation in
  App. \ref{app:snr}).
  The SDSS images are found to be deeper than the HST/ACS-GOODS images
  by 0.5, 0.8, and 0.6 magnitudes in the rest-frame $u$, $g$, and
  $r$ bands respectively.


\begin{table*}
  \caption{Wavelength comparison between different SDSS and HST/ACS bands. The rest-frame wavelengths are calculated assuming a redshift range [0.4-0.6].}
	\begin{tabular}{llcccccc}
	\hline \hline
	\textbf{Images} & & $u$ & $g$ & $r$ & $i$ & $z$ \\ \hline
	SDSS & & 3551\AA & 4686\AA & 6165\AA & 7481\AA & 8931\AA \\ \hline
	& $B$ & $V$ & $i$ & $z$ & & \\ \hline
	HST/ACS & 4312\AA & 5915\AA & 7697\AA & 9103\AA & & \\
	\textit{rest-frame}{\hspace{1mm}}  & [2695-3080]\AA & [3697-4225]\AA & [4811-5498]\AA & [5689-6502]\AA \\ \hline
	
	\end{tabular}
	\label{tab:lambda}
\end{table*}

\begin{table*}
\begin{center}
	\begin{tabular}{l||ccc|ccc}
		& \multicolumn{3}{|c}{SDSS} & \multicolumn{3}{|c}{GOODS (ACS)} \\ \hline \hline
		Telescope diameter $D$ (m) & \multicolumn{3}{|c}{2.5} & \multicolumn{3}{|c}{2.4} \\
		Filter & $u$ & $g$ & $r$ & $V$ & $i$ & $z$ \\
		Exposure time $T$ (s) & 53.9 & 53.9 & 53.9 & 5450.0 & 7028.0 & 18232.0 \\
		Filter effective wavelength $\lambda$ (\AA) & 3551 & 4686 & 6166 & 5915 & 7697 & 9103 \\
		Filter width $\Delta\lambda$ (\AA) & 567 & 1387 & 1373 & 1565,5 & 1017,40 & 1269,10 \\
		Sky surface brightness & 22.15 & 21.85 & 20.85 & 22.74 & 22.72 & 22.36 \\ \hline
	\end{tabular}
\end{center}
	\caption{Observational parameters for both SDSS and GOODS imaging.}
	\label{tab:obs}
\end{table*}

\section{Morphological analysis}

The morphological classification of each galaxy relies on a process
that was formalised in \citet{Delgado10} as a semi-automatic decision
tree, which takes into account the well-known morphologies of local
galaxies that populate the Hubble sequence. To minimize the remaining
subjectivity, three of us (FH, HF, MP) independently classified each
galaxy using this tree. The results were then compared and
disagreement cases discussed until a final classification was agreed
by all classifiers. The decision was based on the examination of the
elements described below : a color map and the decomposition in
several components of the light profile analysis in a single
photometric band ($r$ band for the local sample and $z$ band for the
distant one). We recall that as star formation likely affects the
aspect of a galaxy in the rest-frame blue bands, the morphological
classification must be performed in a (rest-frame) red band in which
the emission is dominated by main sequence stars, which form the most
part of the stellar mass, hence our choice for the working bands.

\subsection{Color images and maps}
Color information is essential to classify the morphology of a galaxy.
A measure of the color in each pixel of the image offers the
possibility to examine different galaxy components/regions separately
and 
assess whether they correspond to simple star forming regions or are
real sub-structures.

To do so, we subtracted pixel by pixel the magnitudes in two observed
bands using a method that provides measurements of both the colors and
the associated signal-to-noise ratios (see \citealt{Zheng05} for more
details). Since all the considered bands have similar
  resolution and sampling, no PSF matching was needed so that the
  images were directly subtracted one from the other. The two bands
used were $u-r$ for the local galaxies and $v-z$ for the distant ones,
ensuring a meaningful comparison between the same rest-frame colors.
For the distant sample we also used the $B-z$ color-map, because the
$v-z$ color did not allow us to classify properly early-type galaxies
at our working redshifts. Indeed, the rest-frame $V$ band corresponds
to the Balmer break region, which can be affected by recent star
formation. Such star formation bursts can occur in the central regions
of a galaxy if some gas was captured 1 Gyr ago, which leads to a blue
core in the $V-z$ color map compared to the rest of the galaxy. We
observed this phenomenon especially in E and S0 galaxies, in which gas
might be more likely prone to infalling motions towards the center,
while in spiral galaxies the gas more likely follow circular
trajectories along the disc. The $B-z$ color map was therefore used to
distinguish between real blue cored galaxies and early-type galaxies
that are only affected by this effect at their centers.

Three-band color images were also constructed using \textit{u-g-r}
bands for the local sample and \textit{V-i-z} bands for the distant
one, as those shown in Fig. \ref{morphseq}.

For the distant sample, we also used infrared $J$ and $H$ HST/WFC3
CANDELS images (see Sect. 2.2.2), which were helpful to remove doubts
when clumpy structures were visible in $z$ band, and distinguish
between irregular galaxies and regular discs: as the infrared bands
are more representative of the galaxy stellar mass, if such structures
were also visible in $J$ and $H$, then they represented a non negligible
part of the galaxy mass, and the galaxy was classified as irregular.

\subsection{Light profile decomposition}
The light profile of each galaxy was analysed in the rest-frame $r$
band, i.e., using the observed $r$ band for the local sample and the
observed $z$ band for the distant sample.

First, the position angle and the axis ratio of the galaxy were
measured using SExtractor. The half-light radius was then derived from
the flux curve-of-growth constructed by measuring the flux in
concentric ellipses of increasing radii. This curve reaches a plateau
when the ellipses covers all the galaxy surface and start sampling the
background. The half-light radius was estimated as the radius at which
the measured flux equals half the total flux (measured on the
curve-of-growth plateau). For both the local and distant
  samples, we checked that the positions of the galaxies in the
  $R_{half}$ vs. $M_{stellar}$ plane are consistent with the relations
  observed in large surveys (see e.g. \citealt{Shen03} and
  \citealt{Dutton11} and references therein).

The two-dimensional surface brightness distribution of each galaxy was
fitted using GALFIT \citep{Peng02} using a linear combination of different
components:
\begin{itemize}
	\item A bulge, modelled by a Sersic \citep{Sersic68} profile
	\begin{equation}
	\Sigma(r)=\Sigma_e e^{-\kappa_n[(r/r_e)^{1/n}-1]},
	\label{sersic}
	\end{equation}
	where $r_e$ is the effective radius enclosing half the total
        light, $\Sigma_e$ is the surface brightness at $r_e$, $n$ is
        the Sersic index, and $\kappa_n$ is a constant depending on
        $n$;
	\item A disc, modelled by an exponential profile 
	\begin{equation}
	\Sigma(r)=\Sigma_0 e^{-r/r_d},
	\label{expdisk}
	\end{equation}
	where $\Sigma_0$ is the central surface brightness, and $r_d$
        is the disc scale-length;
      \item If present a bar was added as an elongated (i.e., with a
        small $b/a$ axis ratio) Sersic component. When a bar was
        present but the resolution not sufficient to resolve it, the
        central Sersic component of the galaxy was considered to be the
        resulting superimposition of the bar and the bulge;
      \item The sky background was modelled as a constant with
          first-guess value estimated from the initial Sextractor analysis
          (see above). This component was let free during the fitting
          process.
\end{itemize}
The ratio between the bulge and the total flux ($B/T$) was estimated
from this decomposition. Note that $B/T$ cannot be straightforwardly
estimated from pre-calculated parameters such as the \emph{frac\_deV}
quantity provided to the SDSS database (see App. \ref{app:fracdev}).

For each galaxy we used an iterative process to find the best
decomposition model. First, a radial surface brightness profile in log
stretch was constructed and carefully inspected for a change in slope,
which evidences at least two components. A first-guess model was then
constructed by fitting visually the two regions of the profile
corresponding to the bulge and the disc, which was used as inputs for
GALFIT. The residual image provided by GALFIT after it has converged
from these initial conditions was visually inspected: if an elongated
and symmetric central structure was identified, then a bar was added
to the model and GALFIT run again using these new initial conditions.
Note that for the smallest galaxies the spatial resolution and
signal-to-noise ratio was not always sufficient to allow GALFIT to
converge with three components, which sometimes resulted in a bar
structure still visible in the residual map.

This iterative method was essential to make GALFIT converge towards
the most meaningful solution. Indeed, the fitting process with GALFIT
is based on a $\chi^2$ minimization using a Levenberg-Marquardt
algorithm. This can efficiently finds a local minimum, which is not
necessarily the global minimum. This makes GALFIT quite sensitive to
the initial conditions and can result in GALFIT converging towards a
solution that has no physical meaning, even if the $\chi^2$ is
minimized to a reasonable value. Degeneracies occur especially when
fitting multi-component models with a large number of free parameters
to faint and distant galaxies that are poorly resolved or have a low
signal-to-noise ratio. Note that in some extreme cases of degeneracies
some parameters had to be constrained to a range of values or even to
be fixed to reduce the parameter space to be explored.

\subsection{Morphological classification}

Taking into account the color maps, the color images, the models
produced by GALFIT, and especially the residual maps (subtraction
between the initial image and the model), each galaxy was classified
into the following morphological classes:

\begin{itemize}
\item Elliptical galaxies (E) must show a prominent bulge with B/T
  ranging between 0.8 and 1, and an overall red color;
\item Lenticular galaxies (S0) must also show a prominent bulge with
  B/T ranging between 0.5 and 0.8; this bulge is required to be redder
  than the underlying disc. The
  disc must be highly symmetric with no signs of regular structure
  such as arms;
\item Spiral galaxies (Sp) must show both a bulge and a disc, or a
  single, pure exponential disc. The B/T ratio must range between 0
  and 0.5, and the bulge, if present, must be redder than the disc.
  The disc must be symmetric but can present regular structures such
  as spiral arms, which can be identified in the residual map
  provided by GALFIT. A central bar can also be present as an
  elongated Sersic profile;
\item Peculiar galaxies (Pec) must show asymmetrical features, which
  can be identified in the residual map. Irregularities associated
  to strong color gradients were assumed to be due to merger events
  (Pec/M). Galaxies with very asymmetrical tidal features were
  associated to merger remnant (Pec/MR). Tadpole-like (Pec/Tad)
  galaxies must show a bright knot associated to an extended tail.
  Irregular galaxies (Pec/Irr) must show only an asymmetric light
  profile, or two off-centered components. Galaxies with a bulge bluer
  than the disc (by more than $\sim$0.2 mag) were also classified as peculiar because of their
  unexpected blue nucleus (Pec/BNG). Galaxies with a half-light radius
  smaller than 1 kpc were classified as compact galaxies (Pec/C).
\end{itemize}

\section{Low surface brightness galaxies}

\subsection{Defining LSB galaxies}

For historical reasons Low Surface Brightness galaxies (LSBs) were
often defined as galaxies dominated by an exponential disc with a
$B$-band central surface brightness fainter than 22 mag.arcsec$^{-2}$.
This limit was first determined from a sample of 36 spiral galaxies
studied by \citet{Freeman70}, adopting a 1$\sigma$ threshold fainter
than the peak of the distribution of their extrapolated disc central
surface brightness. However, this narrow surface brightness
distribution was shown to be due to selection effects
(\citealt{Disney76}) that act against the detection of galaxies
fainter than the sky brightness (see also Sect. \ref{VLSB}). LSB
galaxies were extensively studied in the local Universe selected on
the basis of their blue central surface brightness.

However, because the blue band is very sensitive to star formation,
which was more intense in the past, the blue central surface
brightness distribution of galaxies is found to shift towards brighter
magnitudes with increasing redshift (see, e.g., \citealt{Schade95} who
found a strong evolution of 1.2 mag in the rest-frame B-band central
surface brightness distribution of galaxies at $z>0.5$). It is
therefore important to define LSB galaxies with a criterion that is
not biased by stellar evolution when applying it at different
redshifts. We chose to define LSB galaxies in both the local and
distant samples as galaxies with a disc central surface brightness at
least 1$\sigma$ fainter than the median rest-frame $r$-band central
surface brightness in a given sample.

\subsection{Disc central surface brightness distribution}

In this section, only galaxies with a non negligible disc component
(i.e., S0 and spiral galaxies, as well as peculiar galaxies for which
the light profile reveals a disc component) are considered. The
measurement of the central surface brightness in highly inclined
galaxies is affected by two competing effects: the more inclined a
disc galaxy is, the larger the integrated stellar light is along the
line of sight and the larger dust extinction effects are. To
circumvent this, we restricted the study of LSB galaxies to galaxies
with an axis ratio $b/a\geq0.5$, which corresponds approximately to an
inclination smaller than 60$\degree$. This resulted in sub-samples of
68 local disc galaxies and 78 distant disc galaxies. We checked that
these two sub-samples are still representative of their parent
samples, as evidenced by Kolmogorov-Smirnov tests (which resulted in
probabilities of 97\% and 96\% that the local and distant sub-samples
and the luminosity function at the corresponding redshifts are drawn
from the same parent distributions).

Assuming an infinitely thin disc, integration of Eq.~\ref{expdisk}
gives the total flux of the disc component:
\begin{equation}
	F_{tot}=2\pi r_d^2 \Sigma_0,
	\label{ftot}
\end{equation}
where $r_d$ is the disc scale-length (in arcsec), and $\Sigma_0$ the
central surface brightness as defined by Eq~\ref{expdisk}. By
converting this in logarithmic scale and including the correction for
extinction and cosmological dimming effects, one gets the following
relation to estimate the disc central surface brightness in a given
photometric band:
\begin{equation}
	\mu_0=m_d+2.5\log(2\pi r_d^2)-10\log(1+z)-A_{\lambda},
	\label{mu0}
\end{equation}
where $m_d$ refers to the total magnitude of the disc, $z$ is the
redshift, and $A_{\lambda}$ is the galactic extinction. $m_d$ and
$r_d$ were both estimated by GALFIT when fitting the galaxy profile.
For local galaxies, the galactic extinction, calculated following
\citet{Schlegel98} was taken from the SDSS database. For distant
galaxies, which are observed far from the galactic plane, this
correction can be neglected.

\begin{figure*}
\includegraphics[width=80mm]{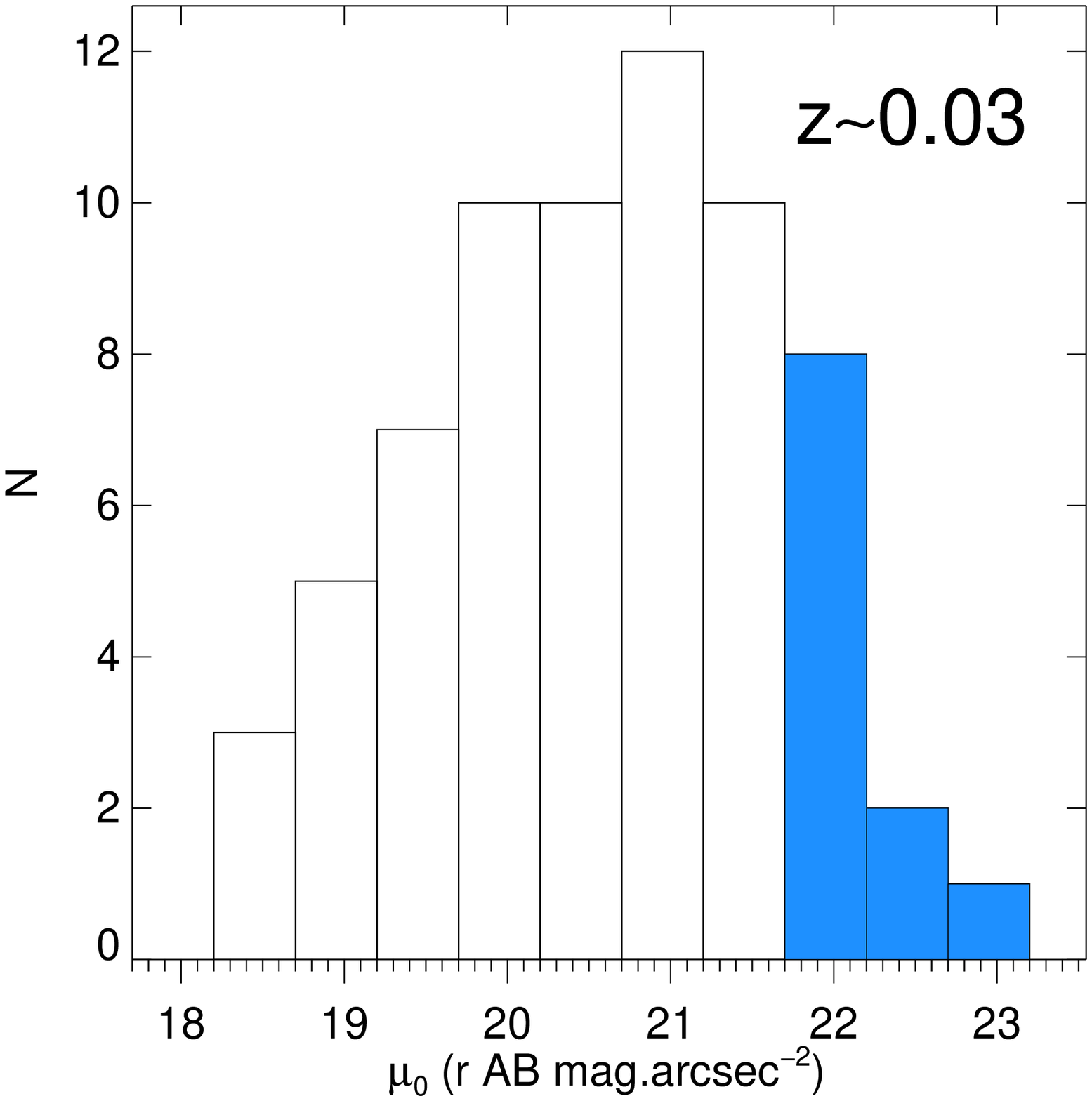} 
\includegraphics[width=80mm]{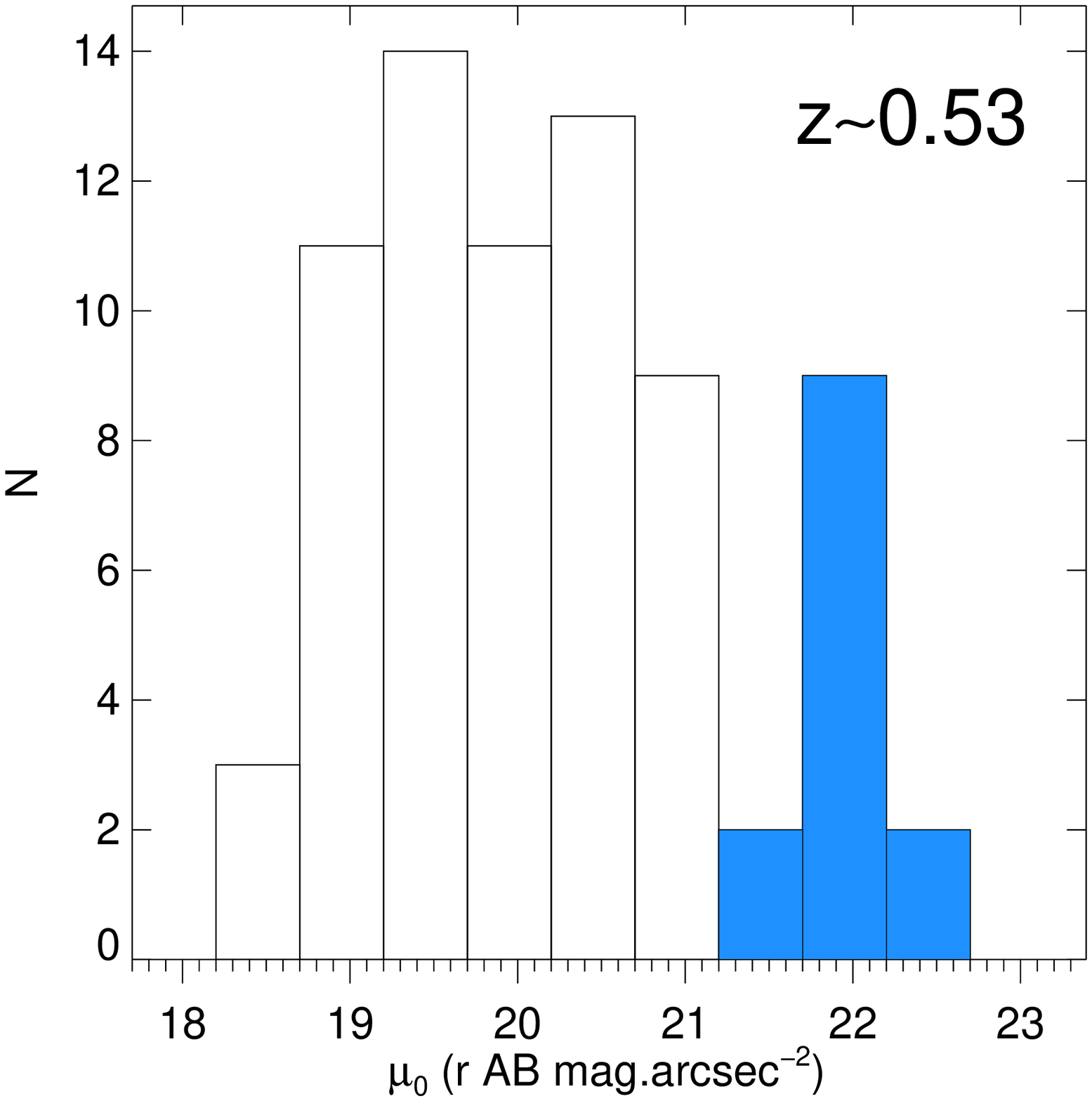}
\caption{Disc central surface brightness distribution for local
  (\textit{left panel}) and distant (\textit{right panel}) galaxies in
  the rest-frame SDSS $r$ band. Median values are 20.7 and 20.2
  mag.arcsec$^{-2}$ for the local and distant samples respectively,
  with a standard deviation of 1 mag.arcsec$^{-2}$ for both. Defining
  LSBs as disc galaxies with a disc central surface brightness at
  least 1$\sigma$ above the median value, the bins corresponding to
  LSBs are shown in blue.}
\label{histmu}
\end{figure*}

The resulting disc central surface brightness distributions in the
local and distant samples are shown in Fig~\ref{histmu}. The median
surface brightnesses in both samples were found to be $\mu_0(r)$=20.7
mag.arcsec$^{-2}$ and 20.2 mag.arcsec$^{-2}$ for the local and distant
samples respectively, with a standard deviation of 1 mag.arcsec$^{-2}$
for both. The median shift of 0.5 mag between the two distributions is
consistent with expectations from a pure passive stellar evolution
over 5 billion years (e.g., \citealt{Charlot96}). As stated above, we
therefore defined LSB galaxies as galaxies with a disc central surface
brightness at least 1$\sigma$ fainter than the above median value,
i.e., $\mu_0(r)\geq$ 21.7 mag.arcsec$^{-2}$ for local galaxies and
$\geq$ 21.2 mag.arcsec$^{-2}$ for distant galaxies.

\subsection{A new class of Giant LSB discs at $z\sim0.5$}

\begin{figure*}
\includegraphics[width=80mm]{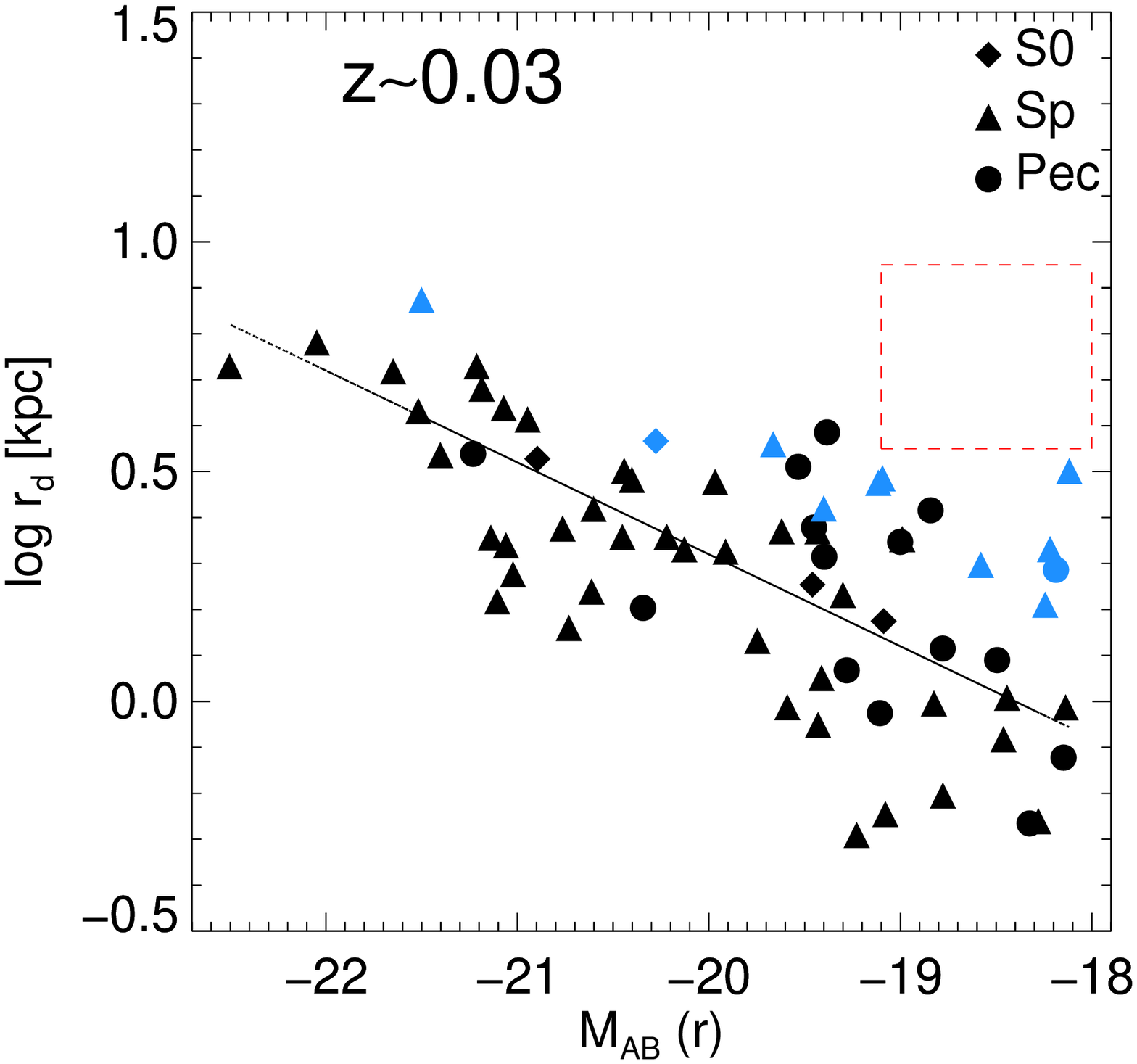} 
\includegraphics[width=80mm]{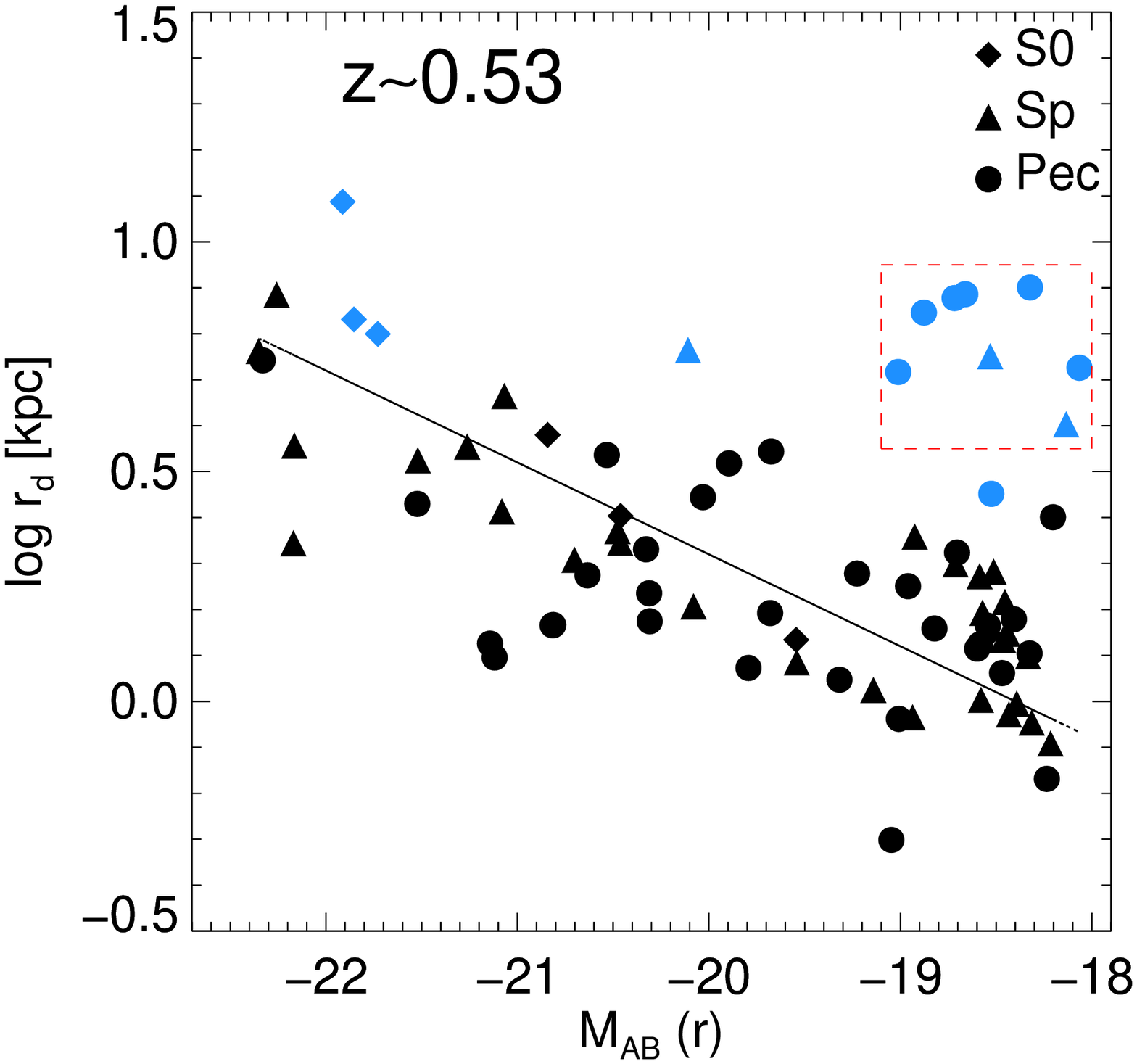}
\caption{Relation between disc scale-lengths $r_d$ and luminosities in
  $r$ band for disc galaxies with $i\leq60\degree$, at low
  (\textit{left}) and high (\textit{right}) redshifts. The plot at
  high redshifts shows a population of giant LSB discs with $r_d\geq
  3-4$ kpc and with low luminosities (see the region ensquared in
  dashed red), which appears not to exist at low redshifts. The black
  line represents how $\mu_0(r)$ is expected to decrease with $m_d$
  from Eq. \ref{mu0} with $\mu_0(r)$=20.7 mag.arcsec$^{-2}$, i.e., the
  median local $\mu_0(r)$).}
\label{Mrrdplane}
\end{figure*}

Figure~\ref{Mrrdplane} shows the disc scale-lengths derived from the
GALFIT 2D fitting as a function of the $r$-band absolute magnitudes
for both the local and distant low-inclination disc galaxies
sub-samples. For both the local and distant samples, the disc
scale-length decreases with absolute magnitude as expected from Eq.
\ref{mu0}, evidencing that the decompositions led to robust disc
components that follow the Freeman's law. By contrast, low surface
brightness galaxies (blue symbols) are clearly shifted upward, i.e.,
they have larger disc scale-lengths compared to high surface
brightness galaxies of similar mass. In the local sample, 80\% of the
LSB galaxies are found to be bulgeless spiral galaxies, half of which
show a bar.

\begin{figure}
\begin{center} 
\includegraphics[width=\linewidth]{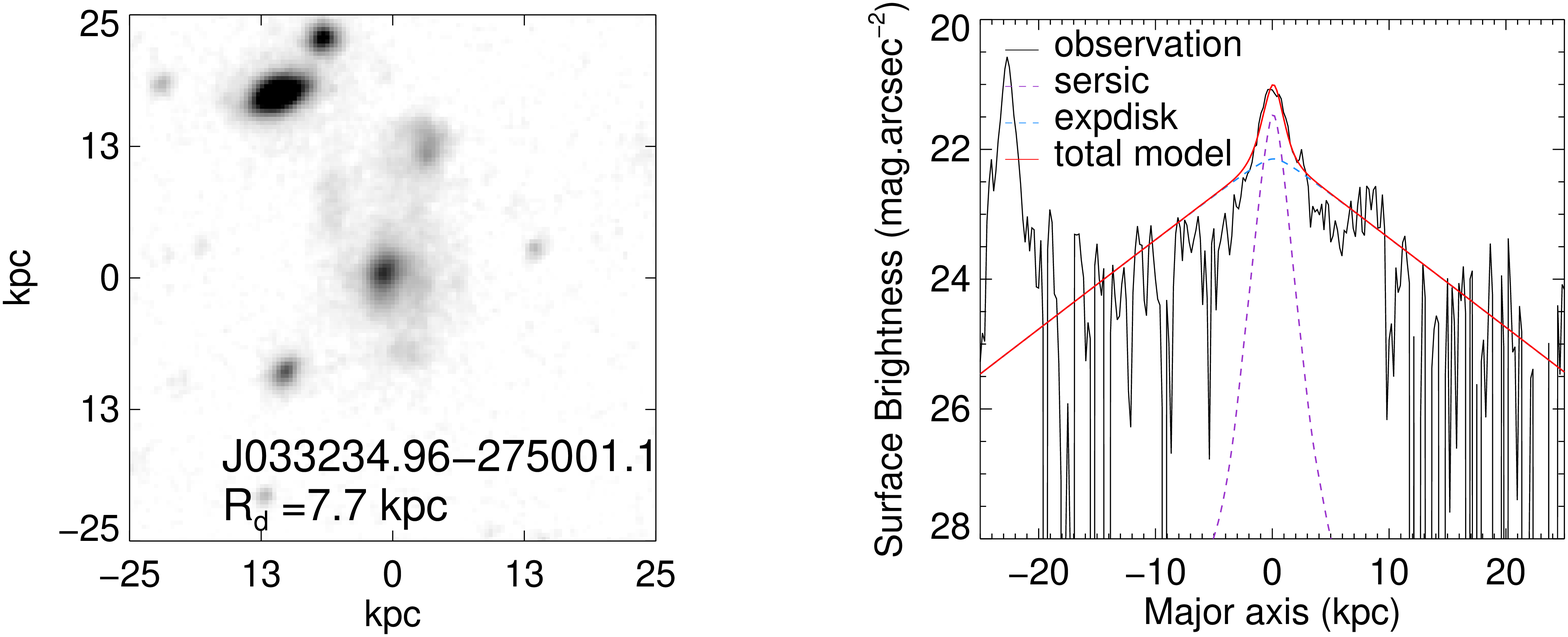}
\includegraphics[width=\linewidth]{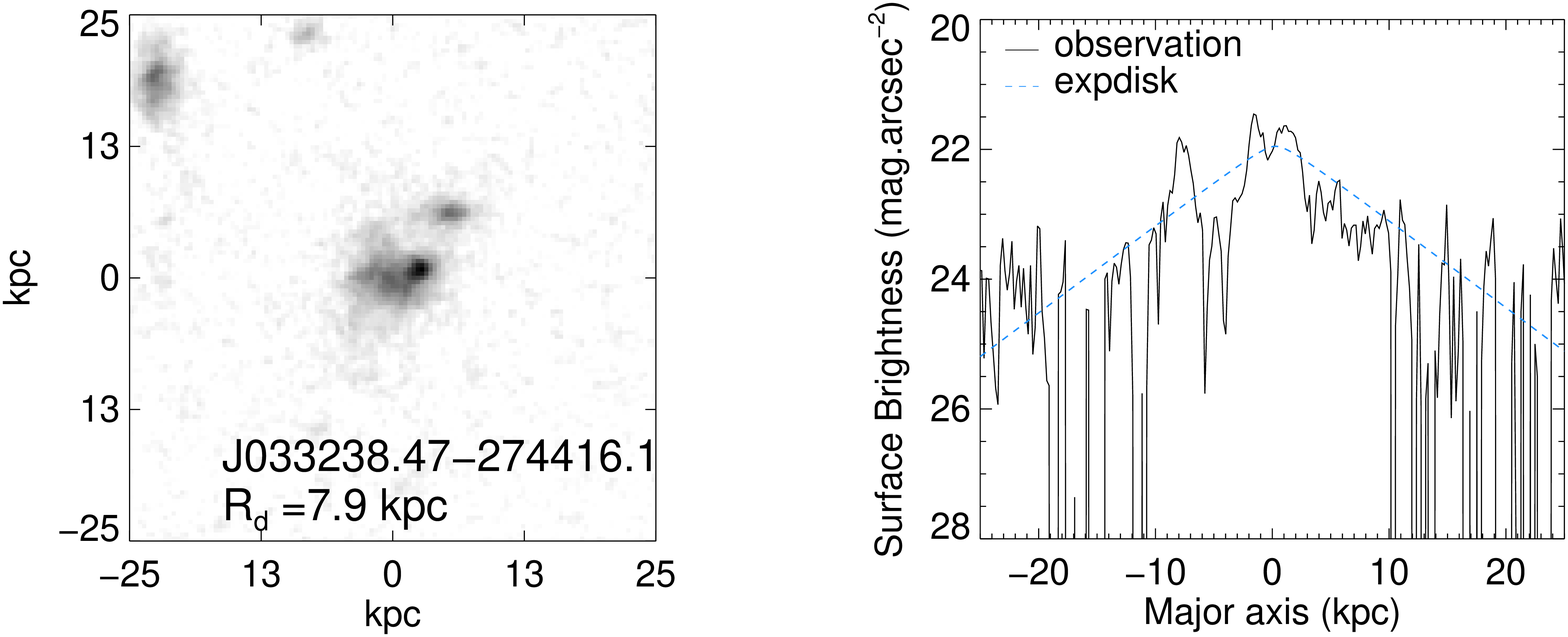}
\includegraphics[width=\linewidth]{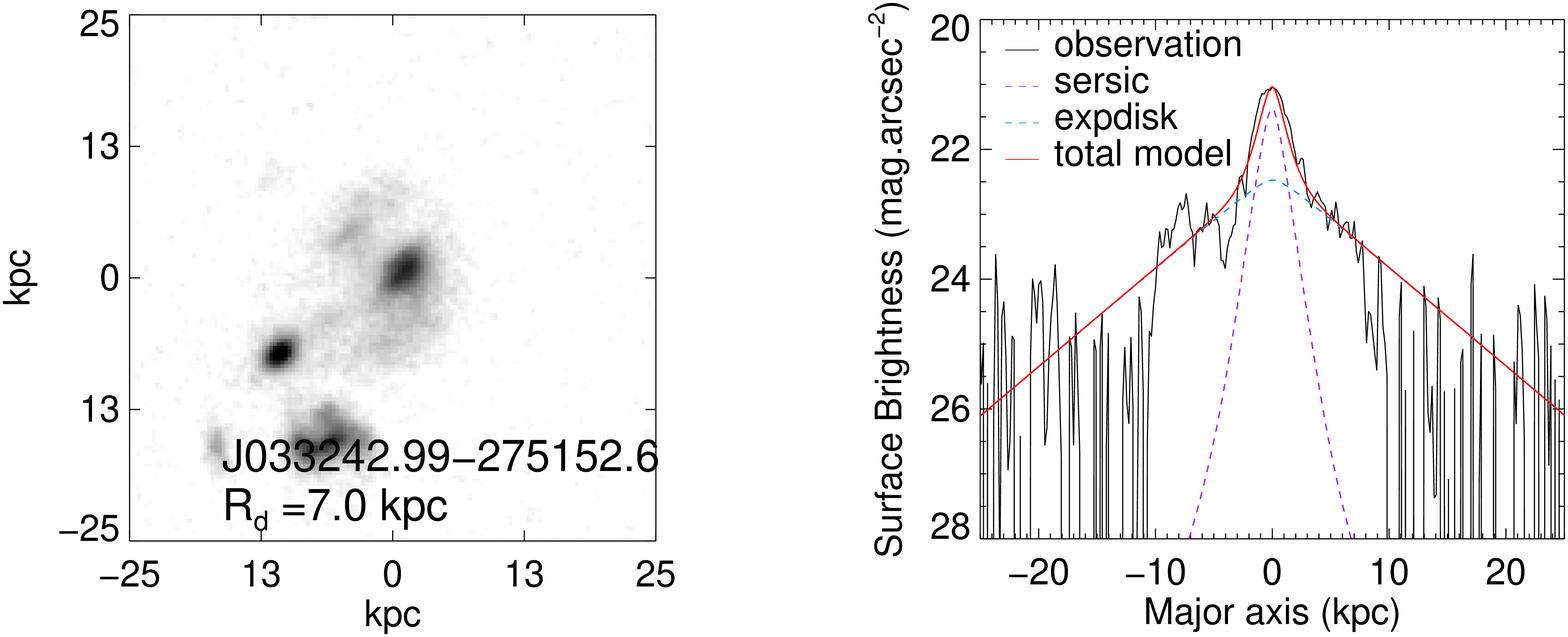}
\includegraphics[width=\linewidth]{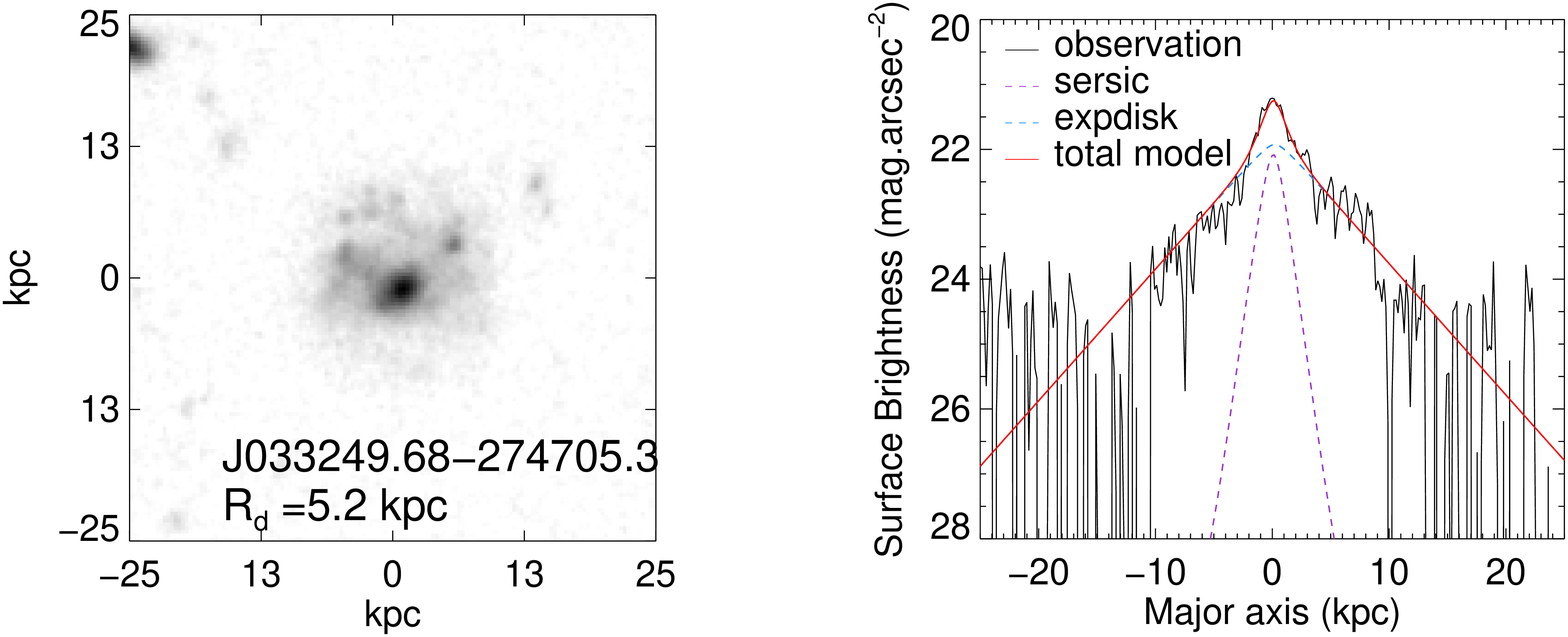}
\end{center}
\caption{Examples of the most extreme LSB galaxies at z $\sim$ 0.5.
  The left panels show the summed $i+z+J+H$ images of the galaxies
  (ACS $i$ and $z$ images were rescaled to the WFC3 resolution). The right panels show the observed
  profile along the major axis of the galaxy disc (black line). The
  profile of the model that has been fitted is superimposed in red.
  When the model comprises several components, the profile of each
  sub-component is shown. Because of the very low signal-to-noise
  ratio of the disc, GALFIT tends to underestimate the disc size. We
  thus kept the disc scale-length parameter fixed during the process :
  we fitted it by visual inspection so that the model profile follows
  the disc wings at best.}
\label{GLSB}
\end{figure}

In the distant sample, 60\% of the LSB galaxies were classified as
Pec/Irr galaxies because they reveal very disturbed morphologies with
asymmetrical brightness profiles and clumps. Their GALFIT model
comprises two components: a Sersic profile with $n \sim 1$ in the
inner component and an extended exponential LSB disc. Comparing the
local and distant samples reveals a population of distant galaxies
with small stellar masses ($M_{AB}(r) >-19$) and very extended LSB
discs ($r_d\geq4-5$ kpc), which appear not to exist at lower
redshifts. The scale-length of these discs are comparable to those of
more massive high surface brightness spiral galaxies, e.g., M31 was
found to have $r_d=5.9\pm0.3$ kpc in $R$ band \citep{Hammer07}. Figure
\ref{GLSB} shows four of these giant LSB discs ($r_d>5$ kpc) found in
galaxies with stellar masses close to that of the LMC. Such extended
LSB discs are further discussed in Sect. \ref{VLSB}.

\section{Building up a morphological sequence of low mass galaxies}

Table~\ref{classresults} summarizes the fraction of E, S0, spiral (Sp)
and peculiar (Pec) galaxies that were found in two mass ranges, i.e.,
$M_{stellar} < 1.5\times10^{10}$ $M_{\odot}$ (low mass or sub-M* range)
and $M_{stellar} \geq 1.5\times10^{10}$ $M_{\odot}$ (high mass range).
This limit corresponds to $M_{AB}(r)\sim -20$.

\subsection{High mass sub-samples and comparison with previous studies}

A fraction of 81-15-5\% of the local massive galaxy population is
found to be Sp-E/S0-Pec galaxies, respectively. At high redshift we
find a slight increase of the fraction of E/S0 and Pec, but a
dramatic decrease of the fraction of Sp. These results can be directly
compared to \citet{Delgado10}, who selected galaxies with $M_J\leq
-20.3$, which also corresponds to $M_{stellar} \geq 1.5\times10^{10}
M_{\odot}$. 
Our
results are found to be consistent, which confirms the evolution of
the fraction of Pec and Sp galaxies with redshift, suggesting that
peculiar galaxies transform into spirals. Nevertheless, we find a less
dramatic quantitative evolution: in our distant high mass galaxy
sample, the fraction of spirals remains higher than the fraction of
peculiar galaxies (45\% and 27\% of spirals and peculiar respectively
instead of 31\% and 52\% in the sample of \citealt{Delgado10}). These differences could
also be explained by the fact that we probe a slightly more recent
epoch of the Universe compared to them (the median redshift of our
distant sample is $z=0.53$ instead of $z=0.65$ in their sample).
Moreover, while they found no evolution in the fraction of E/S0
galaxies, we find a slight increase with redshift, which is not so
significantly when accounting for the relatively large statistical
uncertainties due to the limited sample sizes.  The
consistency between our results and previous works strengthens the
robustness of our classification process.

\subsection{Morphological classification of the low mass subsamples}

We built two morphological sequences of low-mass galaxies at $z\sim0$
and $z\sim0.53$, which are shown in Fig.~\ref{morphseq}. Each stamp in
these figures represents approximately 5\% of the galaxies in the
sample. The counting of bars and LSB galaxies is performed in the
sub-samples of galaxies restricted to $b/a\geq0.5$ because of possible
biases due to inclination and dust on the bar detection as well as on
the central surface brightness calculation (see Sect. 4.2).

\subsubsection{Local low-mass galaxies}

The local low-mass galaxy population consists in 15\% of E/S0, 57\% of
Sp, and 28\% of Pec galaxies (see Tab.~\ref{classresults}). The major
difference with the higher mass galaxy sample is the fraction of peculiar galaxies in the local
Universe, which represents one third of the local low-mass galaxy
population instead of only a few percents in the well-known Hubble
sequence. The fraction of E/S0 is comparable to what was found for
high-mass galaxies. The population of spiral galaxies represents
slightly more than half of the local galaxies compared to more than
75\% for high-mass galaxies. They mainly consists in bulgeless
galaxies (less than 10\% of the spirals have B/T $>$ 0.2). LSB
galaxies, as defined in the previous section, represent about 20\% of
the local galaxy population in this range of mass (and one third of
the spiral galaxies). They are found to be almost exclusively spiral
galaxies, most of them showing a prominent bar.

\subsubsection{Distant low-mass galaxies}

We find that the distant population of low-mass galaxies consists of
5\% of E/S0, 51\% of Sp, and 44\% of Pec galaxies. The fraction of
early-type galaxies is a factor three smaller compared to the local
sample. The fraction of Sp galaxies shows no evolution, in stark
contradiction with high-mass galaxies. They all have
B/T $<$ 0.1, and only 10\% of them that are found to be LSBs. The
fraction of Pec galaxies is larger compared to the local sample. This increase can be
ascribed to the giant and irregular LSB discs that are not detected in
the local Universe (see Sect. 4.3).

\begin{table*}
	\caption{Number of galaxies in each morphological class and
          corresponding fractions for local and distant samples. LSB
          and HSB fractions were calculated in the subsamples of
          low-inclination galaxies (with $b/a\geq 0.5$). Uncertainties given in parentheses were calculated assuming Poisson statistics, as $\sqrt{N}/N$.}

	\begin{tabular}{l|ll|ll||ll|ll}
	& \multicolumn{4}{c||}{$M_{stellar} < 1.5.10^{10} M_{\odot}$} & \multicolumn{4}{c}{$M_{stellar} \geq 1.5.10^{10} M_{\odot}$} \\ \hline
		& \multicolumn{2}{c|}{Local} & \multicolumn{2}{c||}{Distant} & \multicolumn{2}{c|}{Local} & \multicolumn{2}{c}{Distant} \\ \hline 
		
		\textbf{E}  & 11 & (12$\pm$4 \%) & 3 & (3$\pm$2 \%) & 3 & (5$\pm$3 \%) & 6 & (12$\pm$5 \%) \\
		
		
		\textbf{S0} & 2 & (2$\pm$1 \%) & 2 & (2$\pm$1 \%) & 6 & (10$\pm$4 \%) & 8 & (16$\pm$6 \%) \\
		
		
		\textbf{Sp} & 52 & (58$\pm$8 \%) & 50 & (51$\pm$7 \%) & 48 & (81$\pm$12 \%) & 23 & (45$\pm$9 \%) \\ 
		
		\hspace{2mm} \scriptsize HSB &  &\hspace{2mm} \scriptsize 39$\pm$7 \% &  & \hspace{2mm} \scriptsize 44$\pm$7 \% & & & & \\
		\hspace{2mm} \scriptsize\textit LSB & \hspace{2mm} \scriptsize\textit  & \hspace{2mm}\scriptsize\textit 19$\pm$5 \% & \hspace{2mm} \scriptsize\textit  & \hspace{2mm} \scriptsize\textit 7$\pm$3 \% & \hspace{2mm} & & \hspace{2mm} & \\
		
		\textbf{Pec} & 25 & (28$\pm$6 \%) & 44 & (44$\pm$7 \%) & 2 & (4$\pm$2 \%) & 14 & (27$\pm$7 \%) \\
		
		\hspace{2mm} \scriptsize HSB &  & \hspace{2mm} \scriptsize 26$\pm$5 \% &  & \hspace{2mm} \scriptsize 33$\pm$6 \% & & & & \\
		\hspace{2mm} \scriptsize LSB & & \hspace{2mm} \scriptsize 2$\pm$1 \% & & \hspace{2mm} \scriptsize 11$\pm$3 \% & & & & \\ \hline
		
		Total & 90 & (100\%) & 99 & (100\%) & 59 & (100\%) & 51 & (100\%)  \\ \hline

		\textbf{Disc galaxies :} & & & & & & & & \\ 		
		
		\hspace{2mm} \scriptsize without bulge &  & \hspace{2mm} \scriptsize 62$\pm$ 8\% &  & \hspace{2mm} \scriptsize 71$\pm$ 8\% & & \hspace{2mm}\scriptsize 8$\pm$ 4\% & & \hspace{2mm}\scriptsize 21$\pm$ 7\% \\
		\hspace{2mm} \scriptsize with pseudo bulges ($n<2$) & & \hspace{2mm} \scriptsize 16$\pm$ 4\% &  & \hspace{2mm} \scriptsize 19$\pm$ 4\% & & \hspace{2mm}\scriptsize 46$\pm$ 9\% & & \hspace{2mm}\scriptsize 43$\pm$ 9\% \\
		 \hspace{2mm} \scriptsize with classical bulges ($n\geq2$) & & \hspace{2mm} \scriptsize 8$\pm$ 3\% &  & \hspace{2mm} \scriptsize 5$\pm$ 2\% &  & \hspace{2mm}\scriptsize 31$\pm$ 7\% &  & \hspace{2mm}\scriptsize 8$\pm$ 4\% \\ \hline
	\end{tabular}
	\label{classresults}
\end{table*}

\section{Discussion}

\subsection{Potential limitations of the study}
\label{limits}

\subsubsection{Photometric redshifts}

One of the main limitation of the study may be linked to the use of
photometric redshifts to build the distant sample of galaxies. We
indeed first tried to use the ESO-GOODS spectroscopic master
catalogue\footnote{\textit{http://www.eso.org/sci/activities/garching/projects/\\goods/MasterSpectroscopy.html}}
gathering all publicly available redshifts in the CDFS but it did not
allow us to build a complete volume-limited sample down to
$M_{AB}(r)=-18$ at such redshifts. Sampling such small masses is only
currently possible using photometric redshift estimates. The catalogue
of \citet{Dahlen10} provides us with the best photometric redshift
estimates so far, with a scatter in $\Delta z/(1+z_{spec})$ of only
0.04 between the spectroscopic redshifts they retrieved and their
photometric estimates.

\begin{figure}
\includegraphics[width=80mm]{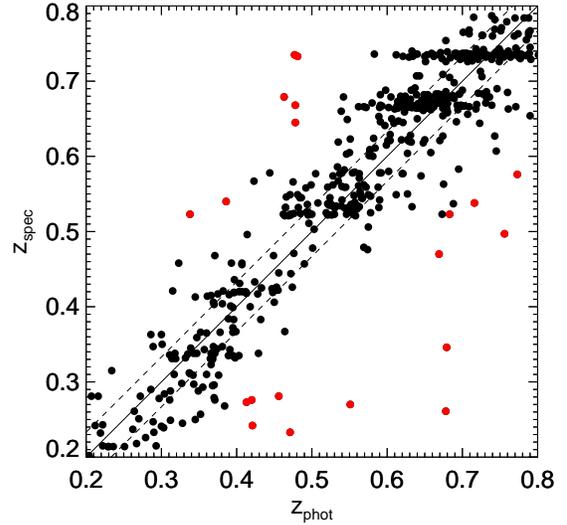}
\caption{Comparison between spectroscopic redshifts (from the ESO
  master spectroscopic catalogue) and photometric redshifts (from
  \citealt{Dahlen10}). Fitting $\Delta z/(1+z_{spec})$ by a Gaussian,
  the scatter is found to be $\sigma=0.033$. Red points represent
  cases of catastrophic failure, defined as $\Delta z/(1+z_{spec}) >
  0.1$.}
\label{crossSpectro}
\end{figure}

We have also cross-correlated the photometric redshift catalogue with
the ESO spectroscopic catalogue to evaluate how errors in redshifts
could affect the sample selection. We have considered only the most secure
redshifts both in the spectroscopic catalogue, in which redshifts are
provided with a quality flag, and in the photometric catalogue from
which we kept only redshifts that were estimated using at least ten
photometric bands. We then fitted the $\Delta z/(1+z_{spec})$
distribution by a Gaussian and found $\sigma=0.033$. The comparison
between spectroscopic and photometric redshifts is shown in
Fig.\ref{crossSpectro}. Since the scatter is much smaller than our
working redshift range, we do not expect possible redshift errors to affect
significantly the sample selection.

Another issue to be considered is the effect of catastrophic failures:
we found that 12\% of objects have $\Delta z/(1+z_{spec}) > 0.1$,
which would represent two stamps in our sequence (Fig.
\ref{morphseq}). Furthermore, Fig. \ref{lumf} shows that the distant
sample follows the same logN-logS behavior that the \cite{Ilbert05}
counts, which are based on spectroscopic redshifts. It indicates that
a similar number of objects near the sample limits (in redshift or in
absolute magnitude) are included/excluded because of photometric
redshifts uncertainties. We conclude that this effect cannot
affect significantly the results.

Errors in photometric redshift estimates could also have an impact on
surface brightness estimates, because of the cosmological dimming
correction that varies as $(1+z)^4$. Indeed, an error of 0.033 in the
redshift estimation leads to an error of 0.10 mag.arcsec$^{-2}$ at
$z=0.5$ in surface brightness. This may affect the results on LSB
galaxies and surface brightness evolution. We were able to retrieve
spectroscopic redshifts for two of the large LSB discs: their
photometric redshifts were found to be 0.462 for both while the
spectroscopy confirms them at 0.503 (J033242.99-275152.6) and at 0.523
(J033225.07-274605.8), respectively. The first (second) galaxy would
have $\mu_0(r)$ = 22.52 (21.80) mag.arcsec$^{-2}$ instead of 22.64
(21.98) mag.arcsec$^{-2}$, implying that their LSB nature remains
unchanged. It illustrates again that the results might be affected by
uncertainties associated to photometric redshifts only in a marginal
way, i.e., only for galaxies near the adopted limits.

\subsubsection{Uncertainties on mass estimates}

\begin{figure*}
\includegraphics[width=80mm]{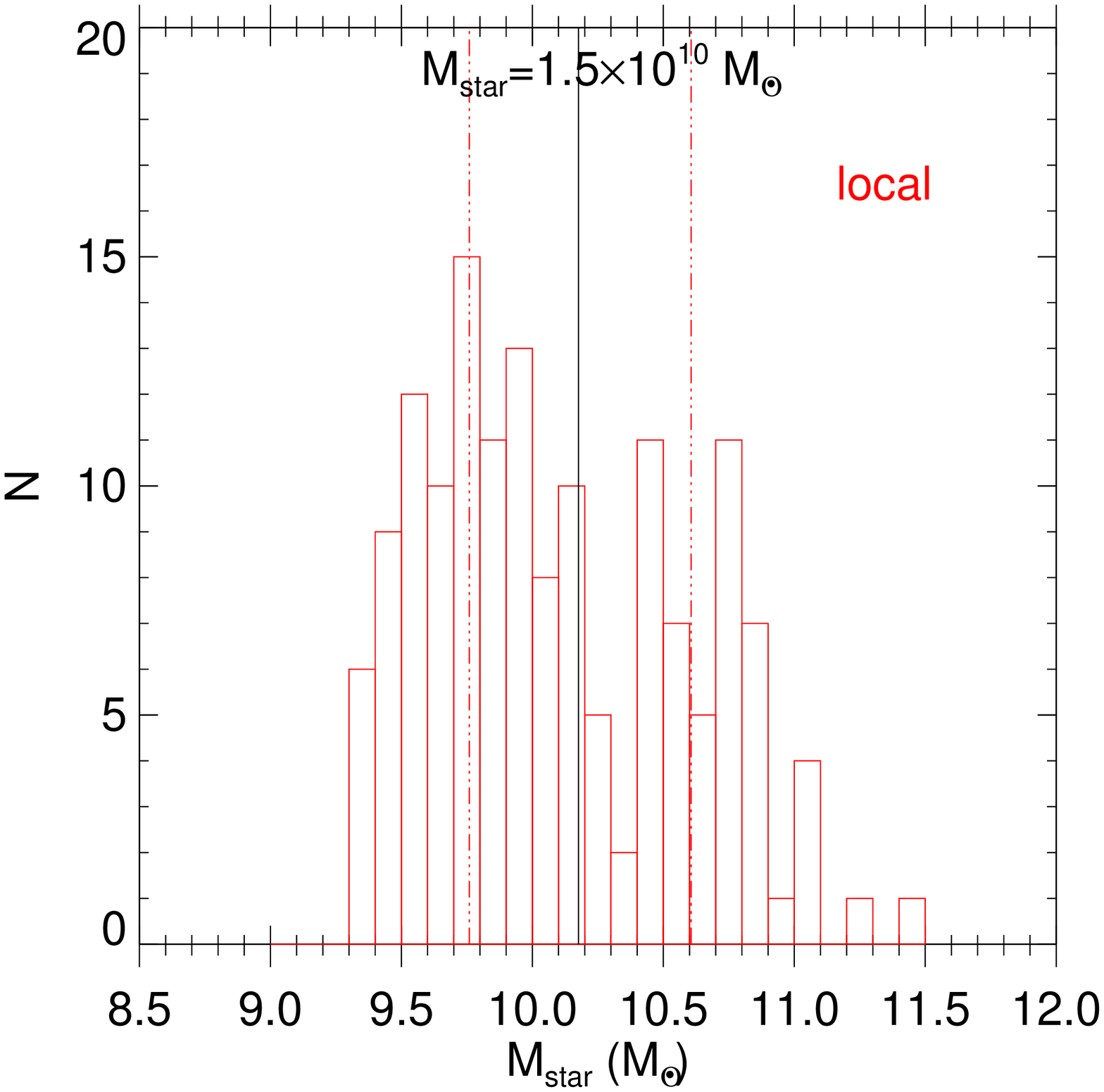}
\includegraphics[width=80mm]{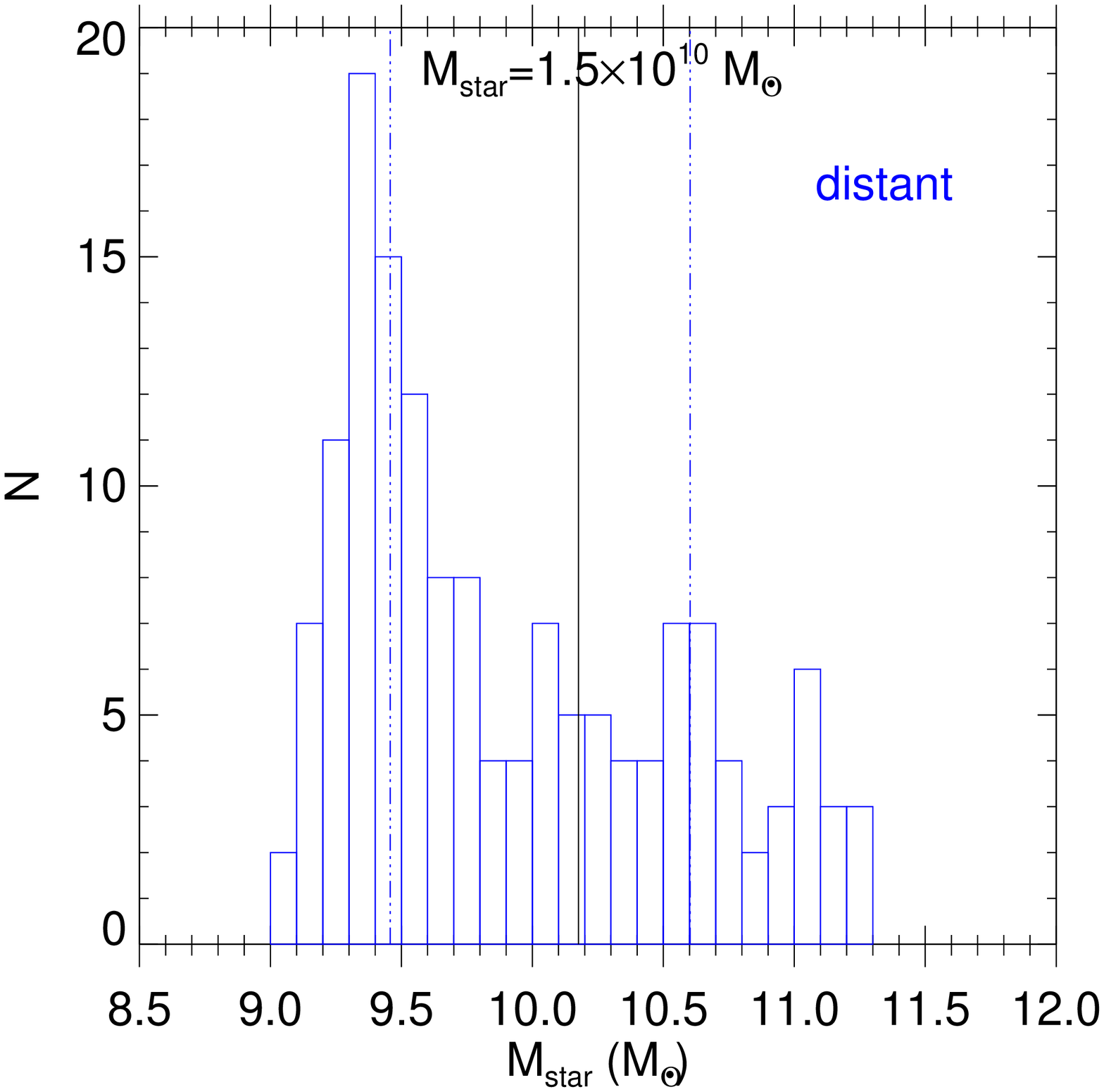}
\caption{Histograms of stellar masses for the local (\textit{left})
  and distant (\textit{right}) samples. Vertical dotted lines
  represent the median mass values for the lower
  ($M_{stellar}<1.5\times10^{10}$ $M_{\odot}$) and upper
  ($M_{stellar}>1.5\times10^{10}$ $M_{\odot}$) mass sub-samples. }
\label{histomass}
\end{figure*}

Mass estimation is still a major issue in extragalactic studies.
Stellar mass estimates are subject to systematic uncertainties related
to the choice of the IMF and the star formation history (see, e.g.,
the discussion in \citealt{Bell03}). Fig. \ref{histomass} evidences a
significant excess of low mass galaxies concentrating near
log($M_{stellar}/M_{\odot}$)=9.5 when comparing the distant and the local
sample. This results from an evolution of the number density of low
mass galaxies during the last 5 Gyr, as evidenced by the comparison
between the local and distant luminosity functions shown in Fig.
\ref{lumf} (see also Introduction). Stellar masses were derived in a
very similar way for both samples and we verified that errors due to
photometric redshifts cannot affect significantly these estimates.

\subsubsection{Could we have missed very low SB galaxies?}
\label{VLSB}
Samples of galaxies with small stellar masses can be affected by the
limits in surface brightness detection. Figure \ref{histmu} (left
panel) shows a significant drop of the local galaxy counts above
$\mu_0(r)\sim22.5$ mag.arcsec$^{-2}$. Accounting for Eq. \ref{mu0} and
the adopted absolute magnitude limit ($M_{AB}(r)<-18$), this suggests
that LSB galaxies with a disc scale-length larger than 2.5 kpc may be
underrepresented/missing in the samples. In particular, comparing the
two panels of Fig. \ref{Mrrdplane}, one can see that there are several
distant LSBs in the interval $M_{AB}(r)=$-18 to -19 with scale-lengths
larger than 3.5 kpc, but none in the local sample (see the red
box in Fig. \ref{Mrrdplane}).

We show in App. \ref{app:snr} that the SDSS images are
  expected to be a factor $\sim$ 1.5-2.6 deeper than the HST/GOODS
  images when considering the $S/N$ obtained per constant 1 $kpc^2$
  apertures and/or per pixels. This suggests that it is unlikely that
  LSB galaxies were missed in the local sample because of an
  insufficient depth compared to the high-z sample. To investigate
  this further, we simulated fake SDSS and HST images of giant LSBs
  and we then tried to recover their morphological parameters with
  GALFIT using exactly the same procedure that the one used for
  observed galaxies (see Sect. 3.2). These simulations reveal that
  extended LSB discs can be well identified in both samples (see
  details in App. \ref{app:snr2}). This strengthens the calculation of
  App. \ref{app:snr} and reveals that possible uncertainties in the
  fitting procedure, e.g., background subtraction, are not severally
  limiting the detection of giant LSB in both samples.

\begin{figure}
\includegraphics[width=8cm]{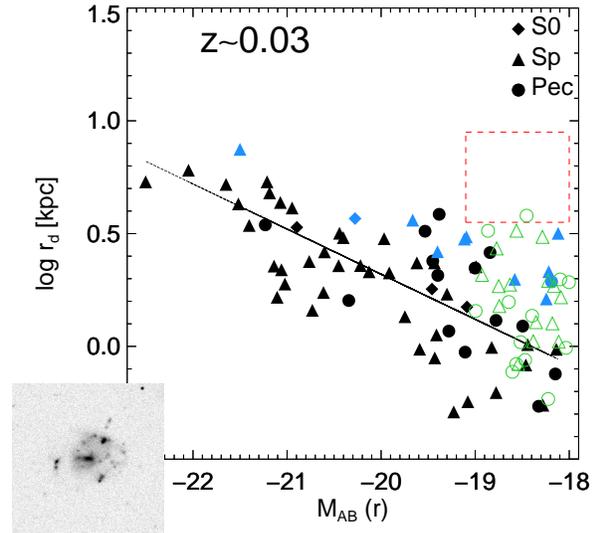} 
\caption{Same as Fig. \ref{Mrrdplane} but including an additional
  sample of 70 galaxies between $-19 < M_{AB}(r) < -18$ shown as open green
  symbols. The stamp image on the lower-left corner shows the summed g+r+i+z image of the only one large LSB found in the additional sample (corresponding to the open green symbol in the red box).}
\label{Mrrdplane_plus}
\end{figure}

Comparing the two panels in Fig. \ref{Mrrdplane}, we find no
  giant LSB among 17 galaxies between $-19 <
  M_{AB}(r) < -18$ in the local sample, while 8 giant LSBs among 35
  such galaxies are found in the distant sample ($\sim$ 23\%, i.e.,
  those in the red box in Fig. \ref{Mrrdplane}). Assuming that the
  fraction of giant LSBs does not evolve with redshift, one would
  expect from the distant sample that the number of local giant LSBs
  is 4. Since the corresponding Poisson uncertainty is $\sqrt{4}=2$,
  the reported evolution of giant LSBs is however only a $\sim$ 2
  $\sigma$ effect within the studied local sample. To check whether
  this can be due to Poisson fluctuations, we selected 70 additional
  SDSS galaxies between $-19 < M_{AB}(r) < -18$ to obtain a sample of
  120 local galaxies in this range of luminosity. We then decomposed
  their light distribution, constructed color maps and images, and
  classified their morphologically, following the method described in
  Sect. 3 and Sect. 4.2 (i.e., we kept only disc-dominated galaxies
  with $b/a > 0.5$ to limit effects due to inclination). These new
  galaxies were added to the $R_d$ vs. $M_{AB}(r)$ plot as open green
  symbols, as show in Fig. \ref{Mrrdplane_plus}. This test reveals
  only one possible giant LSB (with Pec morphology) in the local
  sample. We can therefore conclude that the evolution of the fraction
  of giant LSBs is at least a 4.4 $\sigma$ effect.

 

\subsection{Evolutionary link between the local and distant sequences of low-mass galaxies}

\begin{figure*}
\includegraphics[width=158mm]{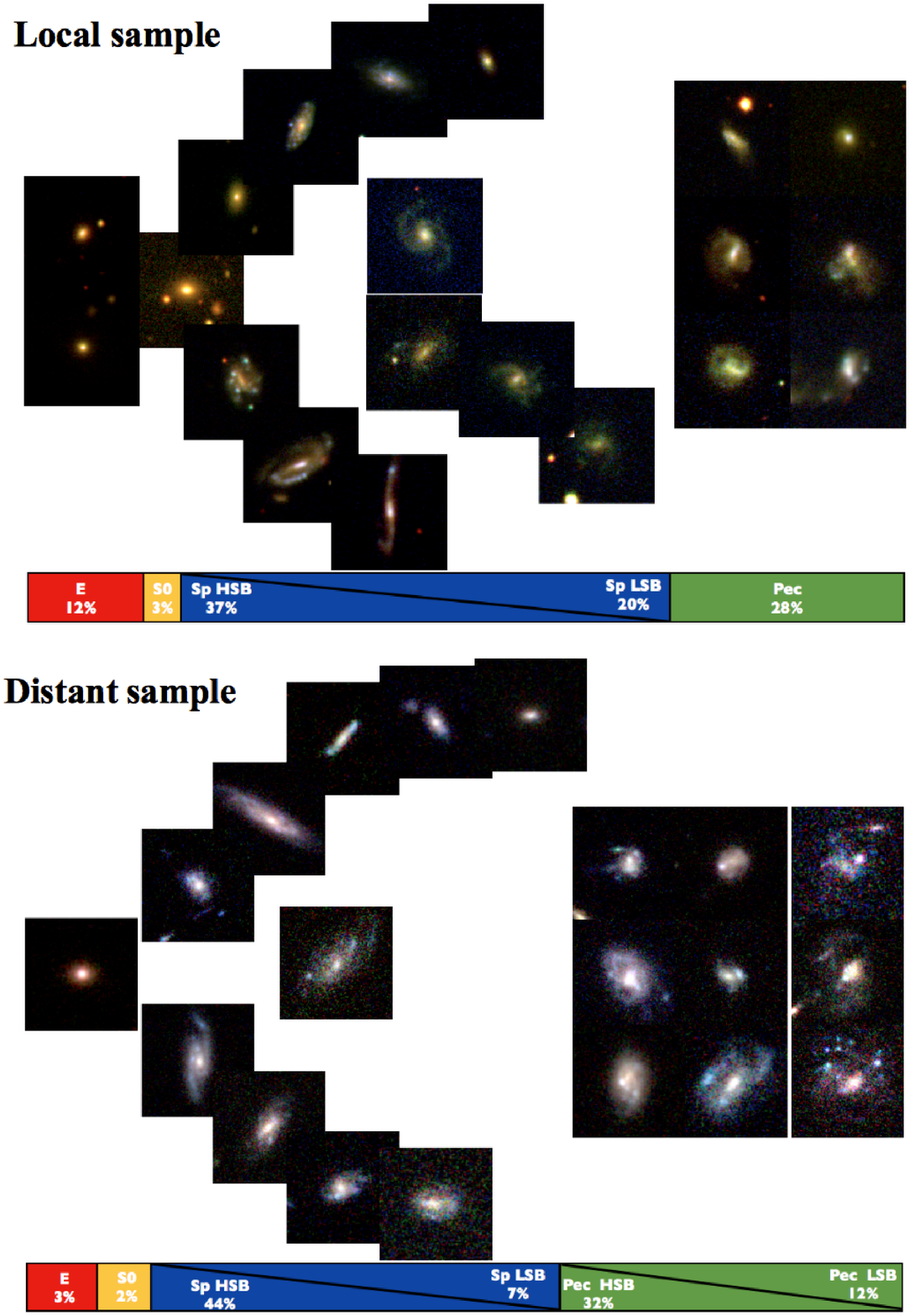} 
\caption{Present-day morphological sequence derived from the local
  sub-M* sample (top panel) and past morphological sequence derived
  from the distant sub-M* sample (bottom panel). Each stamp are 25 kpc
  wide and represents approximatively 5\% of the galaxy population.
}
\label{morphseq}
\end{figure*}

\citet{Delgado10} succeeded to build two Hubble Sequences for
  distant and nearby galaxies and to establish a causal link between
  them. It has revealed a considerable evolution of Sp galaxies,
whose progenitors are mostly Pec galaxies. This is corroborated by
their internal motions revealed by 3D spectroscopy \citep{Neichel08},
and then interpreted and modelled as being the result of gas-rich
major mergers at various phases $\sim$ 6 Gyr ago \citep{Hammer09}.
Within the limitations discussed in Sect. \ref{limits}, Fig.
\ref{morphseq} provides the first ever made attempt to link the
populations of sub-M* galaxies over an elapsed time of $\sim$ 5
Gyr. 

Assuming that morphological peculiarities are associated to merger
events, one can check the consistency of the fraction of Pec galaxies
with the fraction of mergers expected at the two epochs sampled by the
local and distant samples. For this, we used the merger rates derived
by \cite{Puech12} using the semi-empirical model introduced by
\cite{Hopkins10_2}. The fraction of major mergers (defined as stellar
mass ratios $\leq$ 4:1) for events involving galaxies with stellar
mass similar to that of the high-mass samples is expected to be
$\sim$11\% and $\sim$30\% at $z=0$ and $z\sim0.5$ respectively. This
assumes that the morphological classification can identify Pec/major
mergers during a total visibility timescale of 3.2 Gyr. Note that
  parametric morphological classifications can generally detect only
  galaxies during the fusion phase (and partly during the pre-fusion
  phase with a total visibility timescale $<$1.5 Gyr, see, e.g.,
  \citealt{lotz10}), while our morphological classification method
  allows detecting mergers both in the pre-fusion \emph{and} fusion
  phases, which are both found to last $\sim$1.8 Gyr (see Fig. 8 of
  \citealt{Puech12}) for galaxies in this range of mass. We corrected
  the total pre-fusion + fusion timescale for the phases during which
  galaxies would appear as relatively separated pairs with limited
  morphological perturbances, which led us to adopting a total
  visibility timescale $\sim$ 3.2 Gyr. These fractions can naturally
account for the fraction of Pec galaxies found in both the local and
distant high-mass samples (see Tab. \ref{classresults}) with no need
of invoking any other processes.

For sub-M* galaxy mergers, the dynamical timescale for two
  galaxies to merger is expected to be longer by a factor
  $M_{stellar}^{-1/4}$. Since the amplitude of the morphological
  disturbances depends mainly on the mass ratio of the merger, it is
  therefore expected that the total timescale over which major mergers
  between sub-M* galaxies can be identified will be at first order a
  factor $M_{stellar}^{-1/4}$ (i.e., $\sim$80\%) longer compared to
  super-M* mergers with same mass ratios. The fraction of expected
major mergers is then found to be $\sim$8\% and $\sim$31\% in the
local and distant sub-M* samples respectively. There are therefore not
enough major mergers occurring in this range of mass to account for
all Pec low-mass galaxies (28\% and 44\% in the local and distant
samples respectively, see Tab. \ref{classresults}). Accounting for
more minor mergers (i.e., accounting for stellar mass ratios down to
10:1 instead of 4:1) the fraction of galaxies expected to be involved
in major+minor mergers is found to be 13\% and 49\% at $z=0$ and
$z\sim0.5$\footnote{We assumed that the visibility timescales remain
  constant as a function of mass ratio in the range 4:1 to 10:1, as
  found by, e.g., \citealt{lotz10} (see their Fig. 8).}, respectively,
which better accounts for the fraction of low-mass Pec galaxies,
especially at high redshifts. Such an increasing role of minor mergers
in the evolution of sub-M* galaxies is indeed consistent with
expectations from more refined semi-empirical models
\citep{Hopkins10}. It is also consistent with the prevalence of pure
discs and pseudo-bulges amongst low-mass galaxies: 78 and 90\% of
low-mass galaxies are indeed found to have pseudo-bulges (i.e.
  with a Sersic index $n<2$, see \citealt{fisher08}) or to be
bulgeless galaxies in the local and distant samples, respectively
(see Tab. \ref{classresults}).

The 12\% of the low-mass distant galaxies that have extended LSB discs
show strongly perturbed morphologies (see, e.g., Fig. \ref{GLSB}),
which suggest past interactions and contrast with the finding that LSB
discs generally do not show evidence for star formation (e.g.,
\citealt{vanderHulst93}, \citealt{vandenHoek00},
\citealt{Boissier08}). They are also found to be relatively clumpy,
which might also suggest disc instabilities as suggested by
\cite{kassin12}. Interestingly, \cite{guo15} found that minor
  mergers are a viable explanation for the observed redshift evolution
  of the fraction of clumps out to z$\sim$1.5. Their large extent may
share some resemblance with the giant LSB galaxies (GLSB) that were
discovered in the local Universe (\citealt{Sprayberry95} ;
\citealt{Barth07} ; \citealt{Kasparova14}), such as Malin 2. However
the distant LSB discs in this paper have scale-lengths ranging from 3
to 8 kpc, instead of 19 kpc for Malin 2 \citep{Kasparova14}.
\citet{Mapelli08} suggested that GLSB galaxies could form after a very
efficient collision event, which would first produce a ring galaxy
that further evolves into a GLSB galaxy. Such extreme processes are
expected to be very rare because of the required very small impact
parameters. Those are probably not necessary to explain the distant
LSBs found in this paper, though merger/collision events are suggested
from the images. 3D spectroscopy will be necessary to confirm (or
infirm) that major and minor mergers prevail in the evolution of
sub-M* galaxies, including in LSBs.


\section{Conclusion}

Sub-M* galaxies are a very numerous and evolving population, which may
account for a significant part of the star formation density and its
evolution. We have studied in detail two samples of galaxies, which
are representative of the present-day galaxies and their progenitors,
5 Gyr ago. Our analysis is based on a full control of possible biases,
including those related to surface brightness effects. We have
analysed the galaxies 2D luminosity profiles and carefully decomposed
them into bulges, bars, and discs. This was done in a very consistent
way at the two epochs, by relying on the same image quality and
(rest-frame red) filters. Using the morphological decision tree
introduced by \cite{Delgado10}, we have classified them into
elliptical, lenticular, spiral, and peculiar galaxies.

We have discovered that at $z=0$, both massive and sub-M* galaxies
follow similar Hubble Sequences, with a considerable larger number of
peculiar galaxies in sub-M* galaxies. We also found in this range of
mass an emergence of low surface brightness galaxies, which are mostly
found to be disc galaxies. These trends persist in $z\sim0.5$
galaxies, which suggests that sub-M* galaxies have not reached yet a
virialised stage, conversely to their more massive counterparts. The
fraction of distant peculiar galaxies is always high (27-44\%),
consistent with a hierarchical scenario in which minor mergers could
have played a more important role than for more massive galaxies.

We also report the first discovery of a population of giant low
surface brightness galaxies at $z=0.5$, which accounts for 5\% (12\%)
of the overall galaxy (sub-M*) population at z=0.5. These galaxies are
quite enigmatic because their disc scale-lengths are comparable to
that of M31 while their stellar masses are closer to that of the LMC.
Spatially-resolved spectroscopy as well as investigations of the
neutral and ionized gas would be invaluable to establish more firmly
the link between sub-M* galaxies at various epochs and investigate
further the nature of the distant giant LSB discs.

\appendix

\section{S/N of the SDSS and HST images}
\label{app:snr}

In this Appendix, we compare the signal-to-noise ratio ($S/N$)
obtained in the SDSS and HST/ACS-GOODS images. Since the goal is
  to compare the overall morphology of galaxies at different
  redshifts, we first compare the $S/N$ expected in apertures of
  constant physical size. We chose apertures $A$ of size 1 $kpc^2$,
  which correspond to 2.96 $arcsec^2$ at $z_{SDSS}\sim 0.029$ and to
  0.025 $arcsec^2$ at $z_{HST}\sim 0.53$ (see Sect. 2.3). 

The S/N obtained within the area $A$ can be written as follows:
\begin{equation}
  S/N=\frac{N^{source}}{\sqrt{N^{source}+ron^2\times N_{pix}+N^{bkg}}},
\label{snr}
\end{equation}
where $N^{source}$ is the number of detected photons from the source
within $A$, $N^{bkg}$ is the corresponding number of photon from the
background (e.g., sky), $ron$ is the detector read-out noise, and
$N_{pix}$ is the number of detector pixels within $A$. Eq. \ref{snr}
assumes that there are no residual signals from the data reduction.
The galaxies were all observed in background-limited conditions in
all bands except perhaps for the SDSS $u$-band images, for which
galaxies were only on average $\sim$0.25 $mag.arcsec^{-2}$ fainter than
the sky background (see Tab. \ref{tab:snr_regime}). For simplicity, we
assume that all galaxies were observed well in a background-limited
regime, which implies that the above formula can be simplified as:

\begin{table}
\caption{Mean surface brightness of the galaxies within the optical radius $R_{opt} \sim 2\times R_{half}$. Galaxies are significantly fainter than the sky or zodiacal background surface brightness in all bands (except for the SDSS $u$ band, see Tab. \ref{tab:obs}).\label{tab:snr_regime}}
\begin{tabular}{|l|c|c|}\hline
& SDSS & HST/GOODS\\ \hline
Median observed total mag & 16.1 & 22.9 \\ \hline
Median observed $R_{half}$ (arcsec) & 5.0 & 0.5 \\ \hline
$\mu_{opt}$ (mag/arcsec$^2$) & 22.4 & 24.1 \\ \hline
\end{tabular}
\end{table}

\begin{equation}
	S/N \sim \frac{N^{source}}{\sqrt{N^{bkg}}}
\label{snreq}
\end{equation} 

The number of photons from the source detected within $A$ can be
derived from the spectral density of the source surface brightness
$SB_{\nu}^{source}$ (in $erg.s^{-1}.cm^{-2}.Hz^{-1}.arcsec^{-2}$) as
follows:
\begin{equation}
	N^{source}=SB_{\nu}^{source}\times  \Delta\nu \times A \times S \times T_{exp} \times \frac{1}{h\nu_c} \times tr,
\label{sourceph}
\end{equation}
in which $\nu_c$ is the filter central frequency (in $Hz$), $h\nu_c$
represents the energy (in $erg$) per detected photon, the exposure
time is $T_{exp}$ (in $s$), the collecting surface area is $S= \pi
D_{tel}^2/4$ (where $D_{tel}$ is the telescope diameter in $cm$),
$\Delta\nu$ is the filter width (in $Hz$), and $tr$ and the system
global transmission. The source surface brightness spectral density
and the $AB$ surface brightness are simply related by
$\mu_{AB}^{source}=-2.5\times \log_{10}{(SB_{\nu}^{source})}-48.60$ so
that the integrated $AB$ magnitude of the source is $m_{AB}=-2.5\times
\log_{10}{(f_{\nu}^{source})}-48.60$ with $f_{\nu}\sim SB_{\nu} \times
\pi (2.R_{half})^2/4$, in which $R_{half}$ is the source half-light
radius (in $arcsec$) in the considered band.

Similarly, the number of photons from the dominating source of
background light (i.e., the atmospheric sky background for SDSS
images, and the zodiacal light for HST images) detected within $A$ can
be derived from the spectral density of the background surface
brightness $SB_{\nu}^{bkg}$ as follows:
\begin{equation}
	N_{bkg}=SB_{\nu}^{bkg}\times \Delta\nu \times A \times S \times T_{exp} \times \frac{1}{h\nu_c} \times tr,
\label{skyph}
\end{equation}
and then $\mu_{AB}^{bkg}=-2.5\times \log_{10}{(SB_{\nu}^{bkg})}-48.60$
is the $AB$ background surface brightness in magnitude scale.

The $S/N$ obtained within $A$ can be obtained by substituting Eqs.
\ref{sourceph}, and \ref{skyph} in Eq. \ref{snreq}, which leads to:
\begin{multline}
S/N\sim \sqrt{\frac{\pi}{4h}}\times\sqrt{\frac{\Delta\nu}{\nu_c}}\times\sqrt{T_{exp}}\times D_{tel}\times \sqrt{A} \times \\ \sqrt{tr} \times \frac{SB_{\nu}^{source}}{\sqrt{SB_{\nu}^{bkg}}}
\end{multline}

We now compare the $S/N$ that would be obtained if the same source was
observed at $z=0$ and $z=0.53$ using two SDSS and HST/ACS filters. The
ratio between the two $S/N$ is then:
\begin{multline}
	\frac{S/N^{HST}}{S/N^{SDSS}}=\sqrt{\frac{\Delta\nu^{HST}}{\Delta\nu^{SDSS}} \times \frac{\nu^{SDSS}}{\nu^{HST}} \times \frac{T_{exp}^{HST}}{T_{exp}^{SDSS}} \times \frac{tr_{HST}}{tr_{SDSS}}}\\ \times\frac{D_{tel}^{HST}}{D_{tel}^{SDSS}}\times\sqrt{\frac{A^{HST}}{A^{SDSS}}}\times\frac{SB_{\nu}^{source,HST}}{SB_{\nu}^{source,SDSS}}\times\sqrt{\frac{SB_{\nu}^{bkg,SDSS}}{SB_{\nu}^{bkg,HST}}}
\label{snrratio}
\end{multline}

Redshifts and filters were chosen so that the observed HST/ACS filter
for galaxies at $z\sim 0.53$ and the SDSS filter for galaxies at
$z\sim 0$ sample similar rest-frame frequencies (see Sect. 2). For
simplicity, we assume that the match in frequency is perfect so that
the two central frequencies and the two filter widths verify the
following relations:
\begin{eqnarray}
	\nu_c^{HST}=\frac{\nu_c^{SDSS}}{1+z},\\
        \Delta \nu^{HST}=\frac{\Delta \nu^{SDSS}}{1+z}.
\end{eqnarray}
Finally, one has to account for cosmological dimming that decreases
the source flux as $(1+z)^4$ and energy conservation (i.e.,
$SB_{\nu}\times\nu_c=$ cste) that implies
$SB_{\nu}^{HST}=SB_{\nu}^{SDSS}\times(1+z)$, which both lead to:
\begin{equation}
	SB_{\nu}^{HST}=\frac{SB_{\nu}^{SDSS}}{(1+z_{HST})^3}.
\end{equation}
For simplicity, we assume here that $z_{SDSS}=0$ for the local galaxy sample.

Injecting these last three equations into Eq. \ref{snrratio} yields the following relation: 
\begin{multline}
	\frac{S/N^{HST}}{S/N^{SDSS}}=\sqrt{\frac{T_{exp}^{HST}}{T_{exp}^{SDSS}}\times\frac{tr_{HST}}{tr_{SDSS}}}\times\frac{D_{tel}^{HST}}{D_{tel}^{SDSS}}\times\sqrt{\frac{A^{HST}}{A^{SDSS}}} \\ \times\frac{1}{(1+z_{HST})^3}\times\sqrt{\frac{SB_{\nu}^{bkg,SDSS}}{SB_{\nu}^{bkg,HST}}},
\label{snfinal}
\end{multline}
with $z_{HST}=0.53$. Filters transmission curves were retrieved from
the STScI\footnote{http://www.stsci.edu/hst/acs/analysis/throughputs}
and
SDSS\footnote{http://classic.sdss.org/dr7/instruments/imager/index.html\#filters}
websites. The system transmission was assumed to be the value of the
transmission curve at the filter effective wavelength. Other relevant
parameters are listed in Tab. \ref{tab:obs} or detailed in Sect 2. 


Using Eq. \ref{snfinal}, the $u$, $v$, and $i$ band SDSS images are
found to be $\sim$ 1.5, 2.2, and 1.7 deeper than the $v$, $i$, and $z$
band HST/GOODS images, respectively (i.e., by factors of 0.5, 0.8, and
0.6 in magnitude scale). Typical uncertainties on these ratios are
$\sim \pm$ 0.1 mainly ascribed to background variations in the SDSS
images. Note that this largely results from the $A^{HST}/A^{SDSS}$
term in Eq. \ref{snfinal}, since GOODS/HST images are instead found to
be $\sim$ 2 magnitudes deeper than the SDSS images if constant 1
arcsec$^2$ apertures are considered. If now consider the expected
  S/N per pixels (with $\Delta _{pix}^{HST}=0.03$ arcsec and
  $\Delta _{pix}^{HST}=0.396$ arcsec), which is more relevant
  for characterizing the effect of S/N on the morphological GALFIT
  decomposition, the $u$, $v$, and $i$ band SDSS images are found to
  be $\sim$ 1.9, 2.6, and 2.1 deeper than the $v$, $i$, and $z$ band
  HST/GOODS images, respectively (i.e., by factors of 0.7, 1.1, and
  0.8 in magnitude scale).

\section{Testing the sensitivity of GOODS and SDSS images}
\label{app:snr2}

There are very detailed studies of the accuracy with which GALFIT can
recover bulge and disc parameters from HST images in the literature
(see, e.g., \citealt{Haussler07}). As expected, these studies show
that this accuracy decreases as a function of the surface brightness
of the objects. In this Appendix, we use simulated galaxies to
complement these studies in the particular regime of the giant LSBs
with both extended \emph{and} faint discs. For this, we simulated a
grid of 36 fake giant LSBs, which were modelled using two bidimensional
exponential disc profiles, one corresponding to the underlying disc of
the galaxy and another one corresponding to the bulge (since all
bulges in the giant LSBs were found to be well fitted by a Sersic
component with $n=1$). The disc scale length and central surface
brightness of the bulge were fixed to the mean values of the bulges
fitted in the distant giant LSBs (0.91 kpc and 20.84 mag/arcsec$^2$,
respectively), while the structural parameters describing the
exponential discs were chosen to sample the range of disc scale-lengths
and central surface brightnesses observed in the distant giant LSB
galaxies, i.e., with $r_d=[4,5,6,7,8,6.29]$ kpc and
$\mu_0=[21.8,22.0,22.2,22.4,22.6,22.15]$ mag/arcsec$^2$ (with the last
value corresponding to the mean disc scale-length and central surface
brightness of the distant giant LSB discs).

\begin{figure}
\centering
\includegraphics[width=0.45\linewidth]{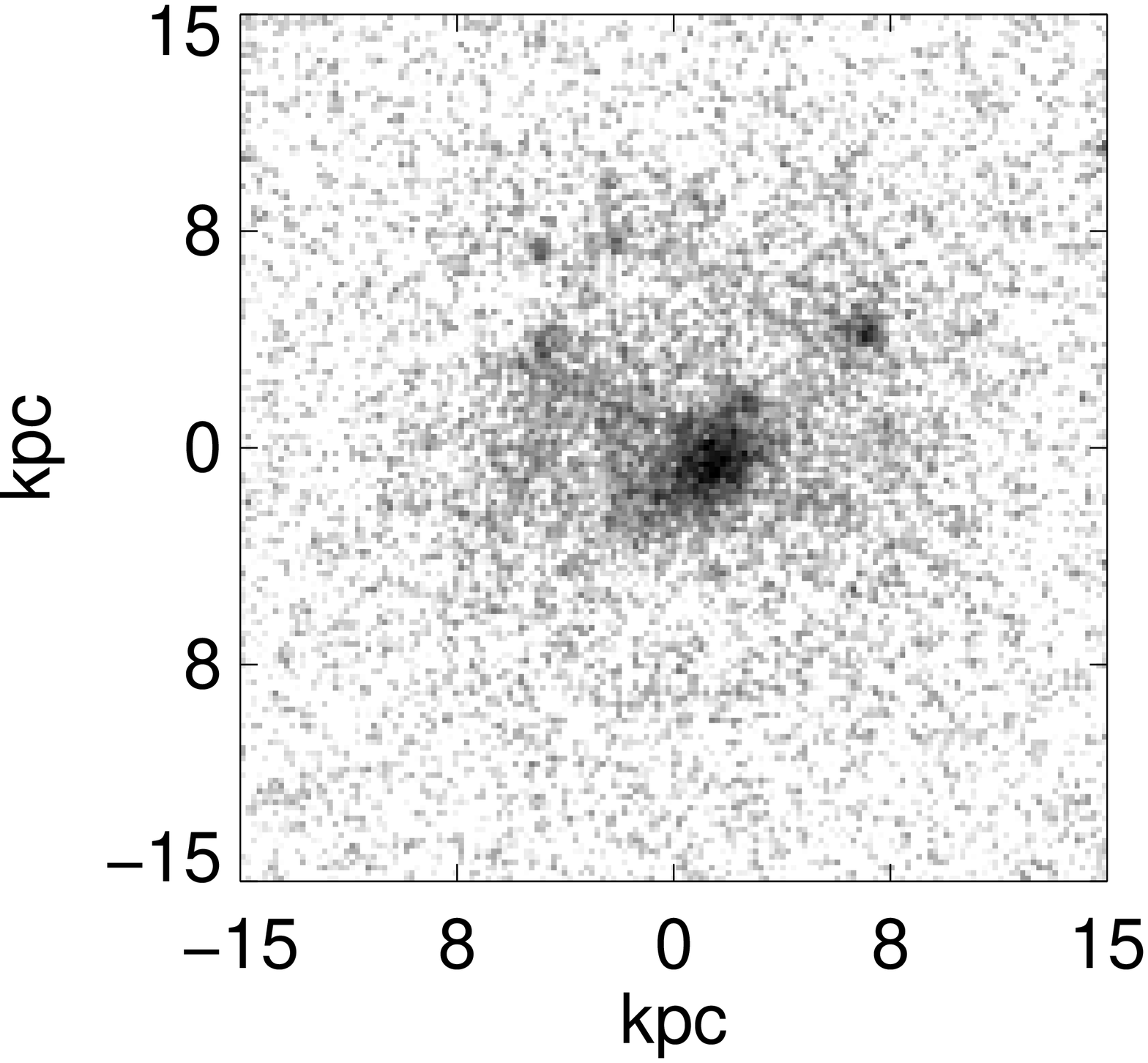}
\includegraphics[width=0.45\linewidth]{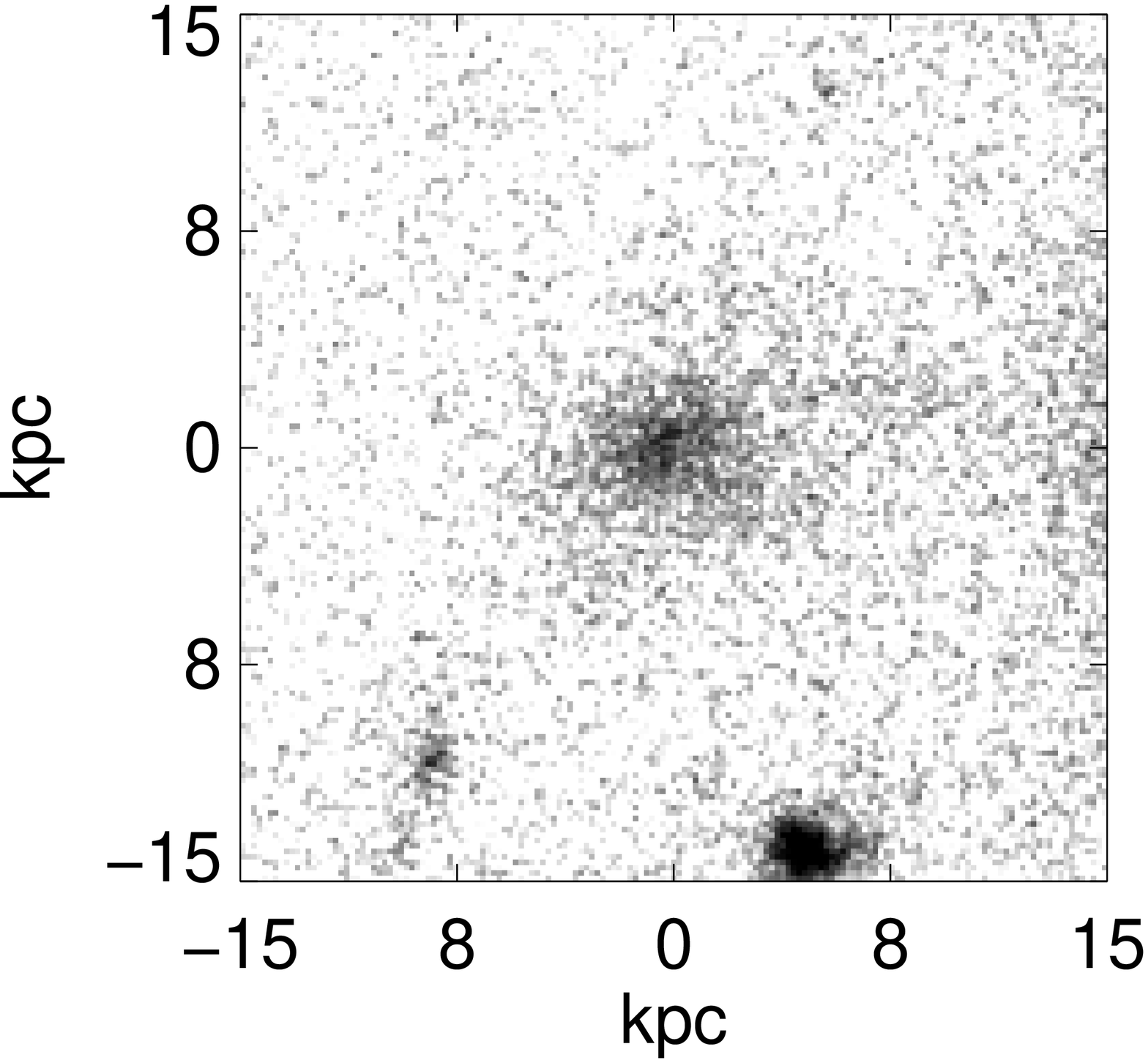} \\
\includegraphics[width=0.45\linewidth]{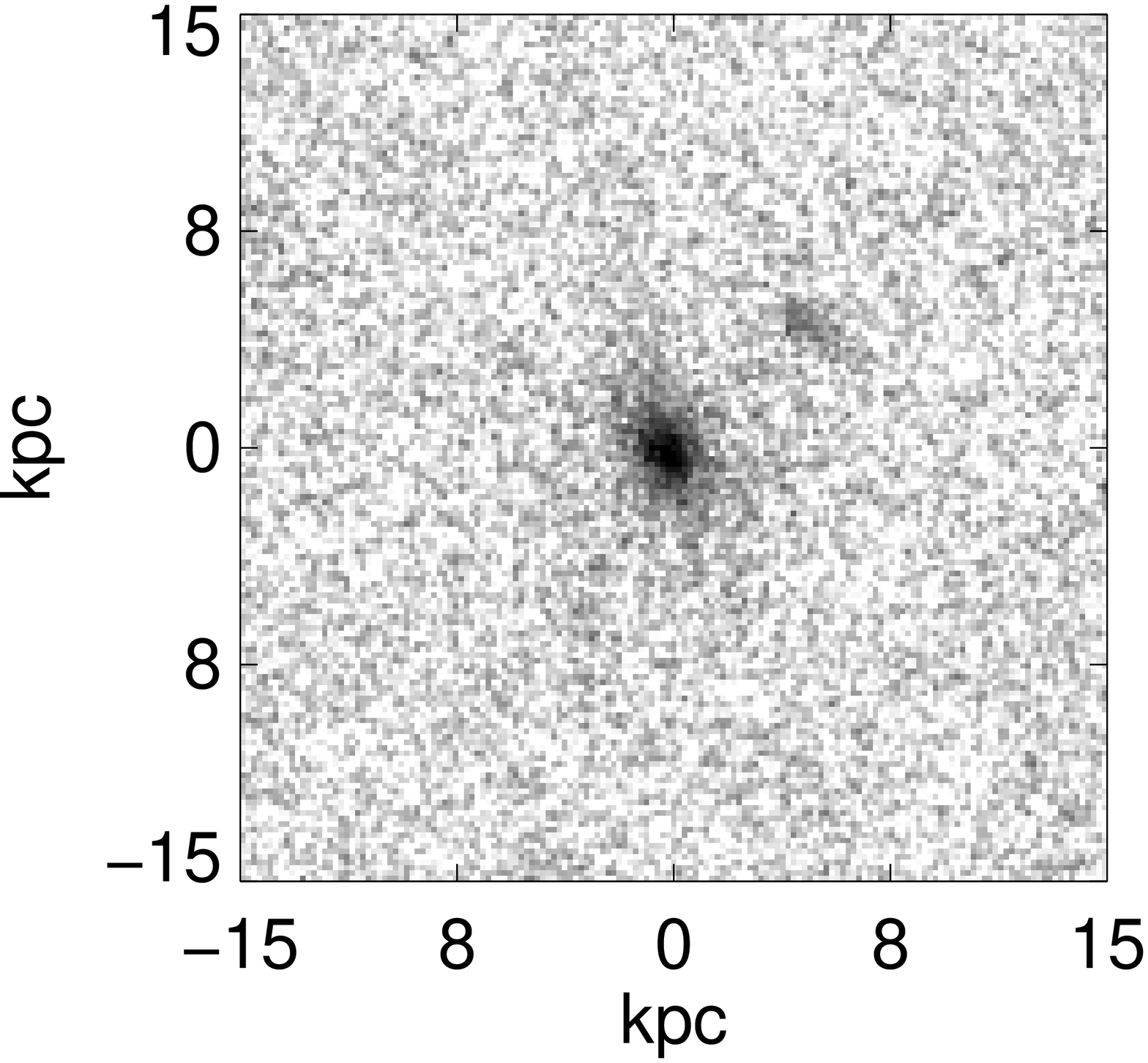}
\includegraphics[width=0.45\linewidth]{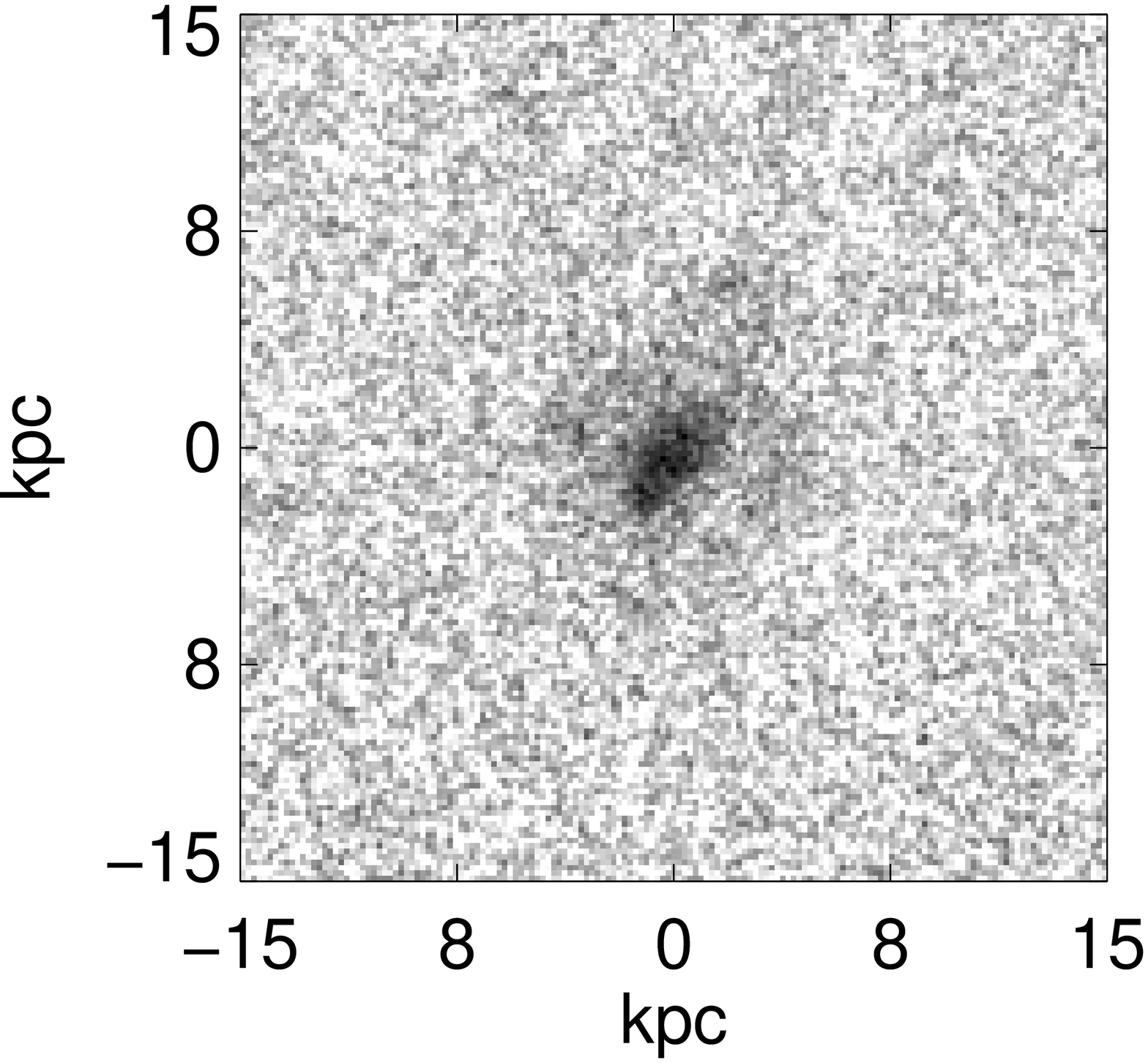} \\
\includegraphics[width=0.45\linewidth]{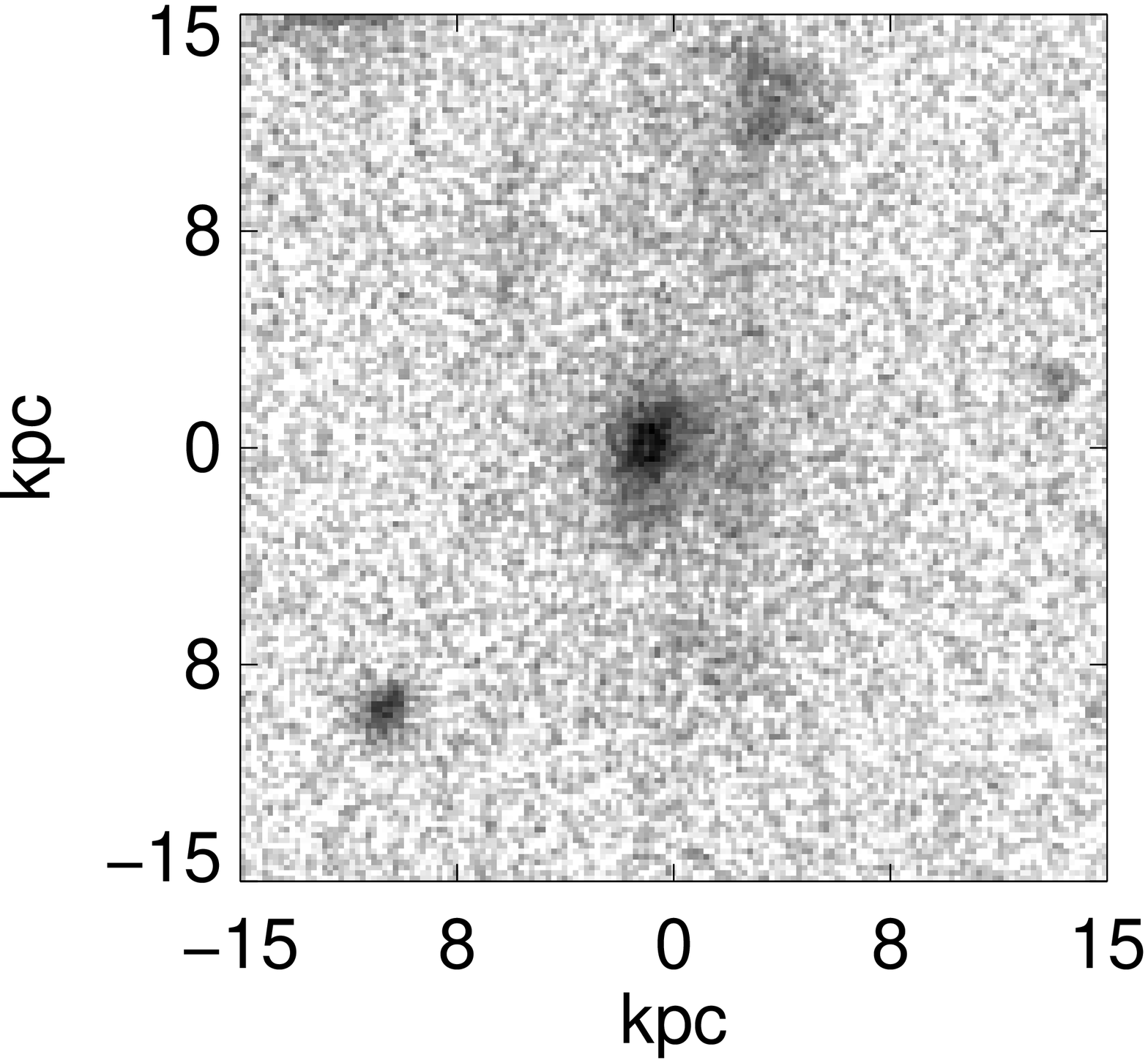}
\includegraphics[width=0.45\linewidth]{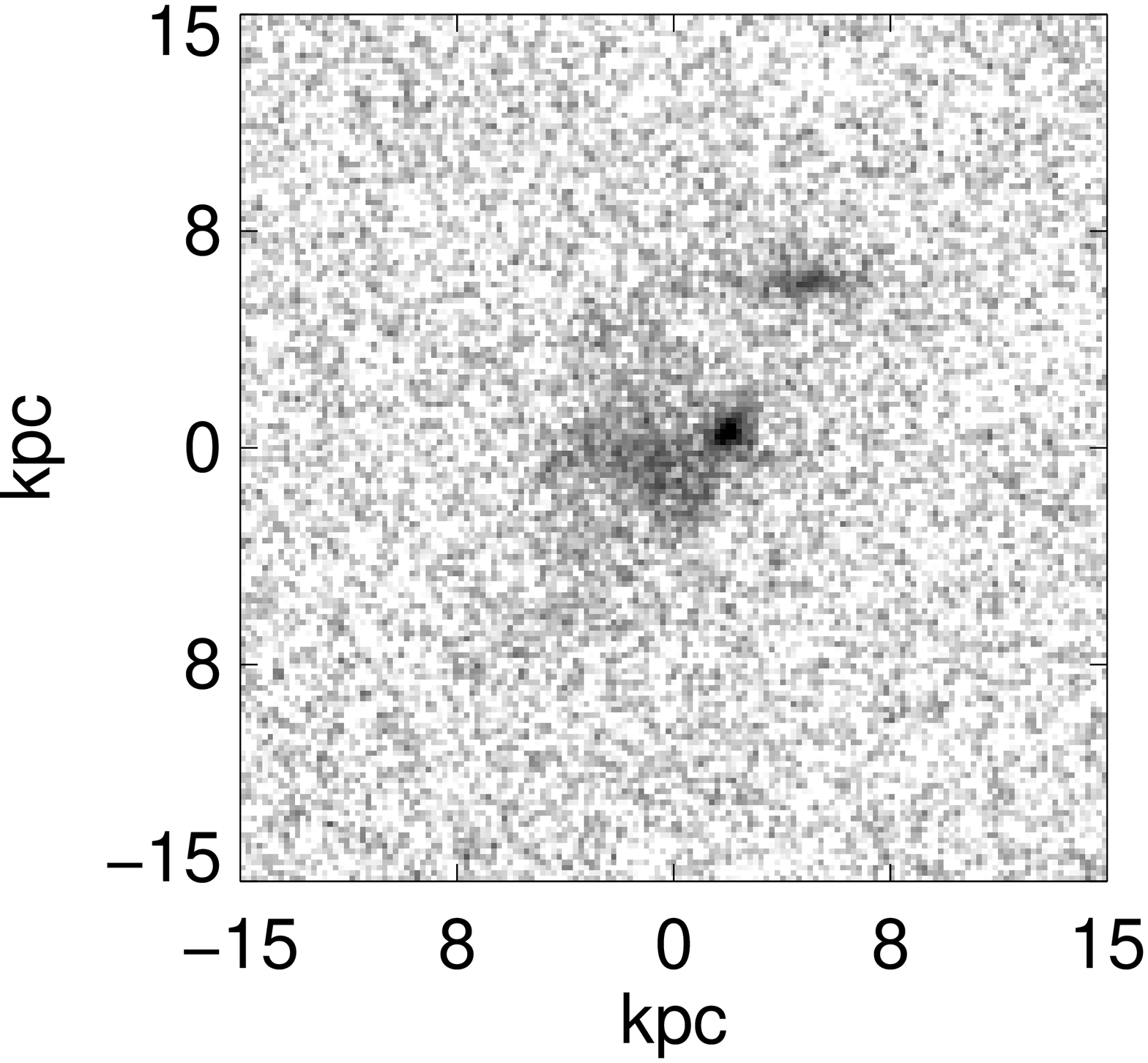} \\
\includegraphics[width=0.45\linewidth]{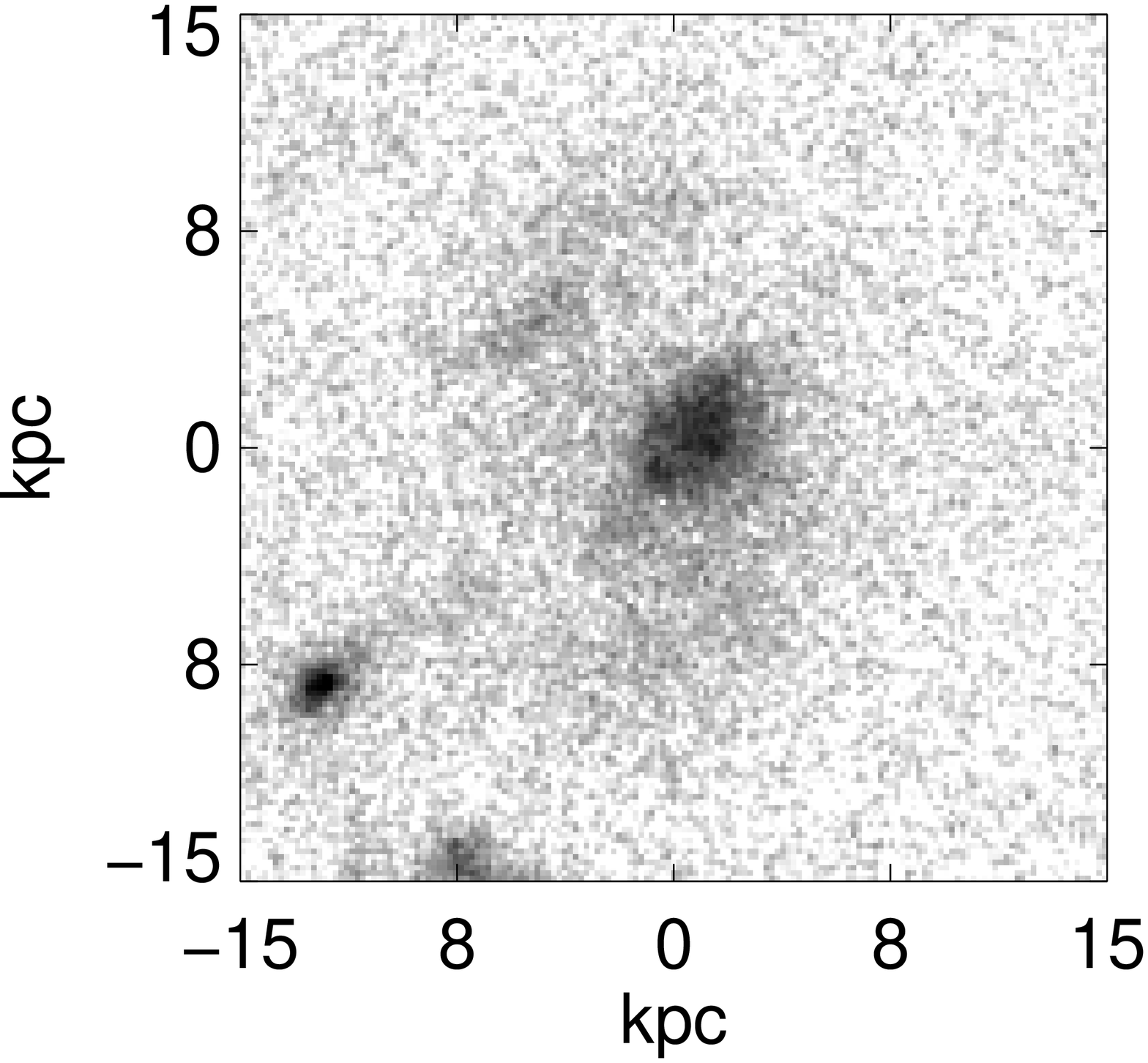}
\includegraphics[width=0.45\linewidth]{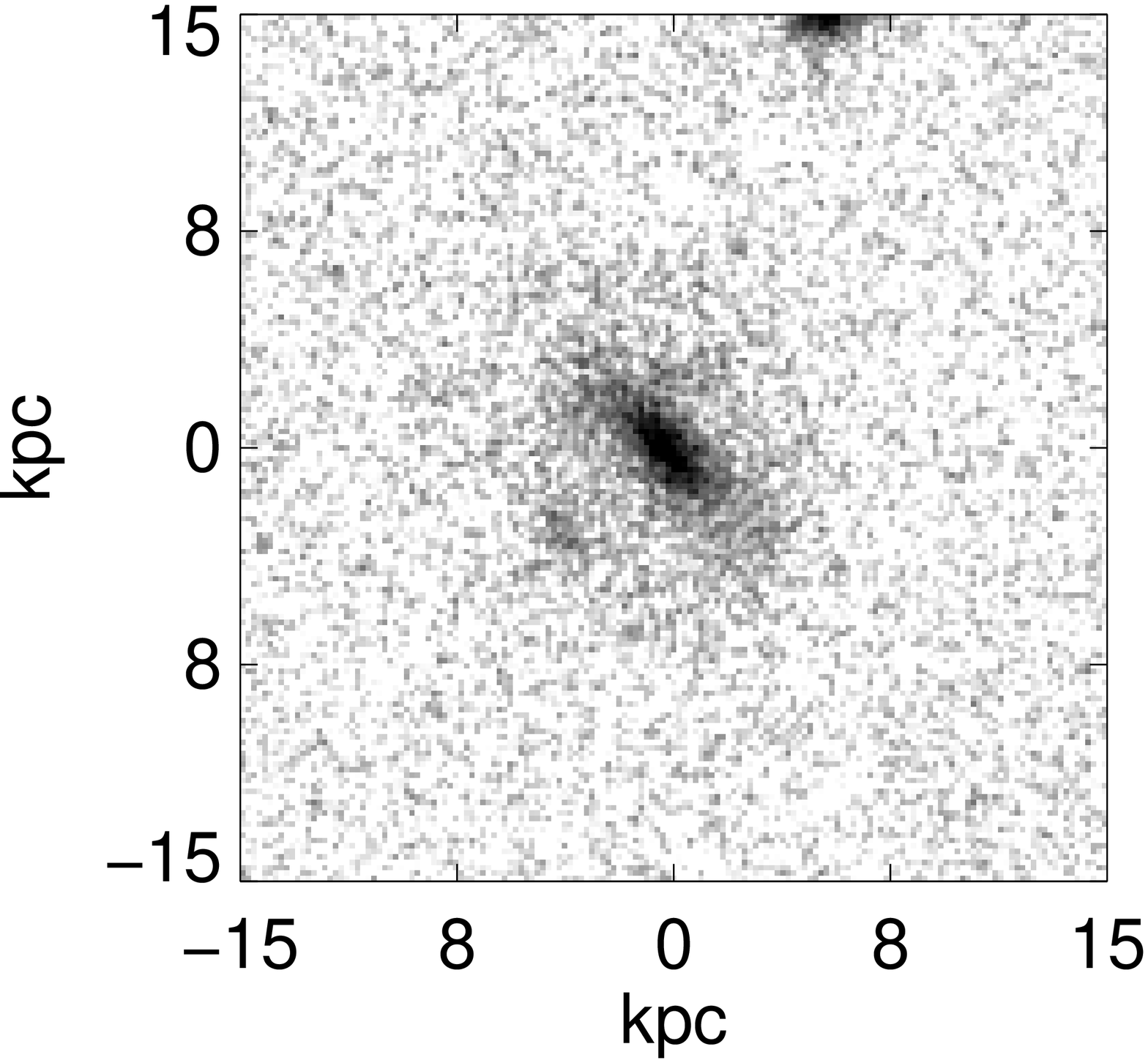} \\
\includegraphics[width=0.45\linewidth]{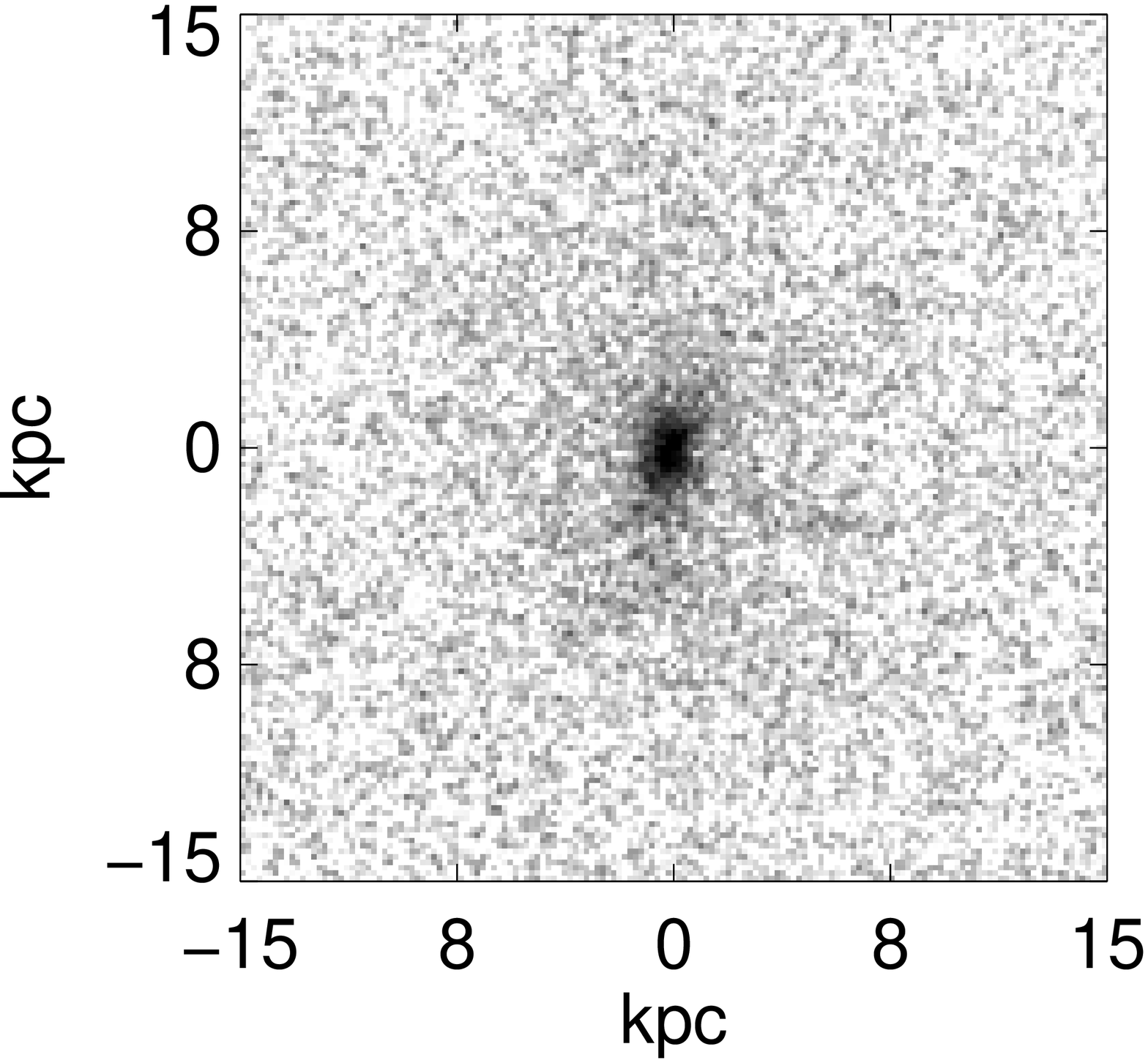}
\caption{Images in $z$ band of the 8 distant giant LSB galaxies. For
  comparison, the last image at the bottom shows the simulated distant
  (GOODS) giant LSB.}
\label{Fig:lsb_fake_comparison}
\end{figure}

For each case two images were created with a pixel size such that
$\Delta pix=FHWM/10$, where $FWHM$ is the full width at half maximum
of the PSF, i.e. 1.4 arcsec for SDSS and 0.1 arcsec for GOODS. Doing
so, the images were oversampled by a factor $\sim$3 compared to real
SDSS and GOODS images. The images were then convolved with a PSF,
assumed to be a simple Gaussian with the same $FWHM$ as in the
observed images, and then rebinned to match the pixel scale of the
real survey images (0.396 arcsec/pix for SDSS and 0.03 arcsec/pix for
GOODS), while conserving the total flux. We cut real sky frames from
empty regions in the SDSS and GOODS images, which were added to the
simulated images to accurately account for real background
fluctuations. We show in Fig. \ref{Fig:lsb_fake_comparison} the
observed distant giant LSBs together with a simulated distant giant
LSB corresponding to the mean disc values.

\begin{figure}
\centering
\includegraphics[width=\linewidth]{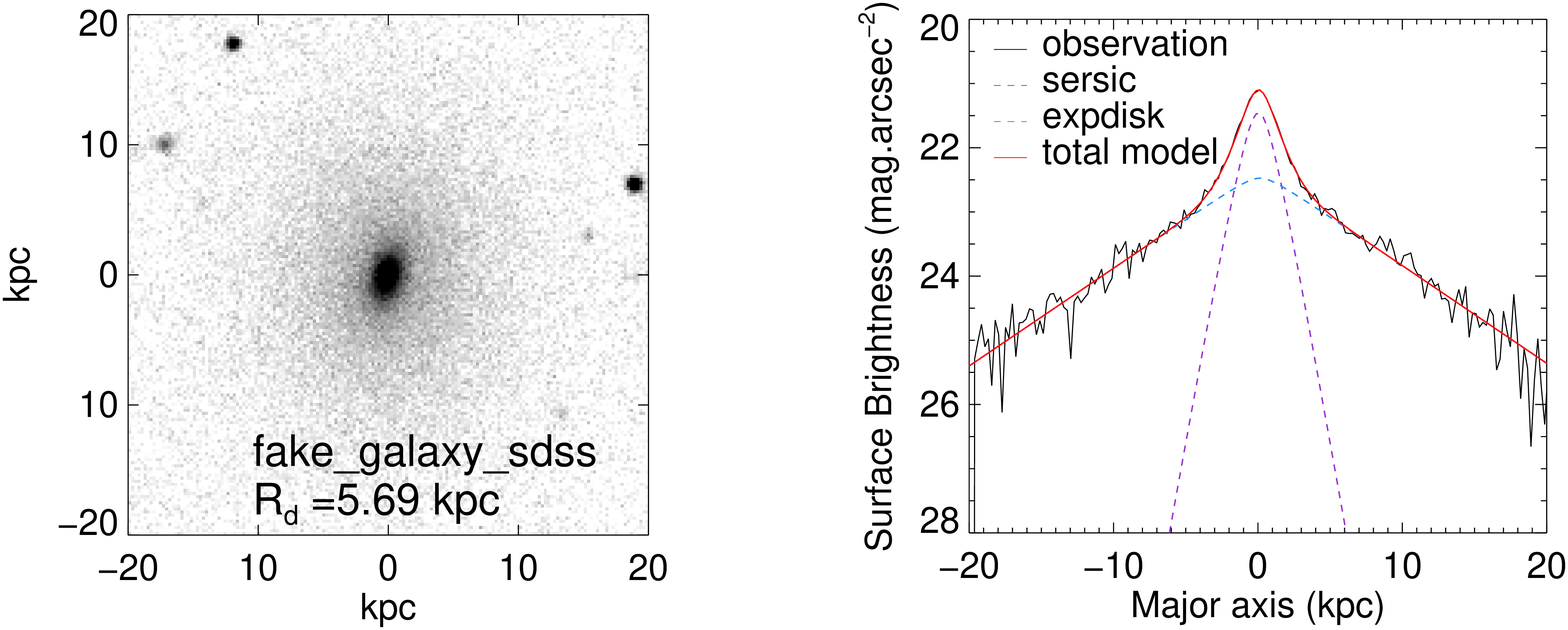} 
\includegraphics[width=\linewidth]{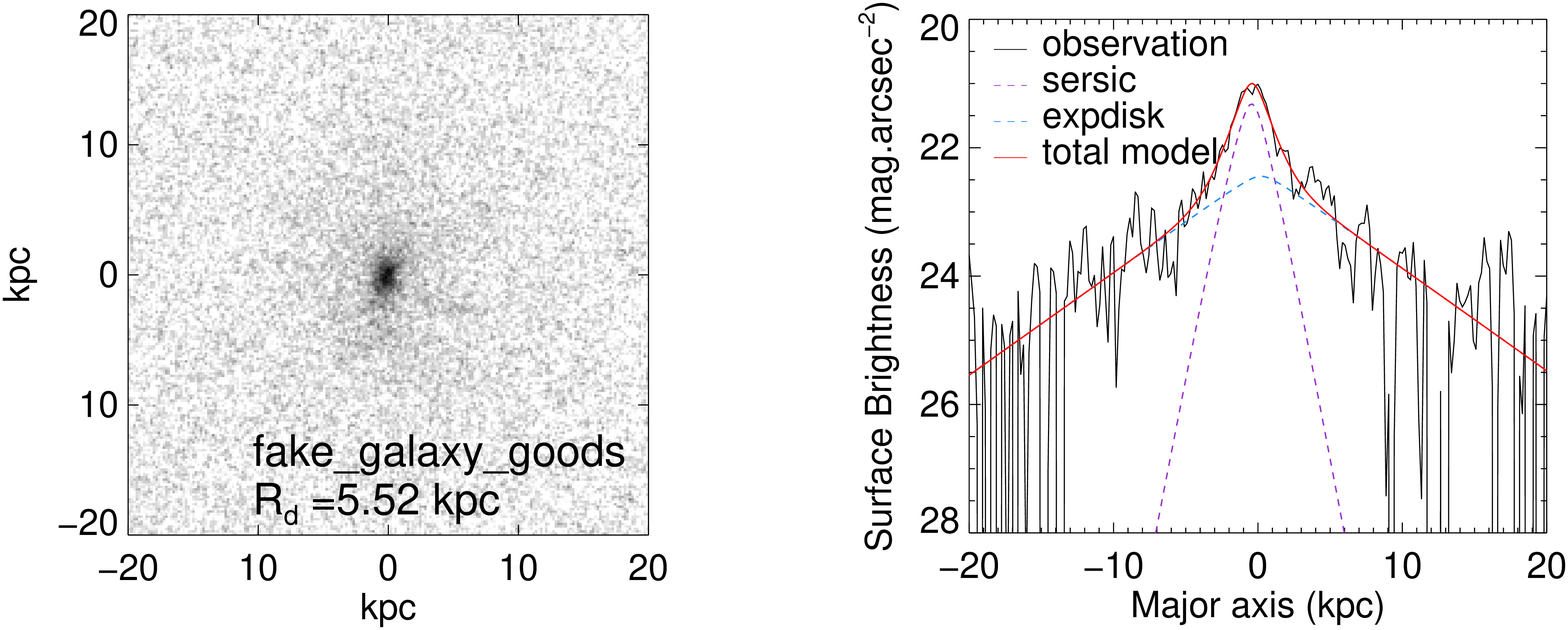} 
\caption{Simulated giant LSB images and light profiles for the SDSS
  (upper panel) and GOODS (lower panel) surveys. The right panels show
  the light profiles along the major axis over which the profile of the
  model that has been fitted has been
  superimposed.\label{Fig:fake_galfit} }
\end{figure}

The light profile of the 36 local and 36 distant simulated giant LSBs
was then decomposed using GALFIT and analysed using the method
described in Sect. 3.2. The mean and r.m.s. values of the differences
between the initial and fitted values of the bulge and disc $r_d$ and
$\mu_0$ were derived, as listed in Tab. \ref{Tab:fake_galfit}.
Fig.~\ref{Fig:fake_galfit} shows the results corresponding to the case
sampling the mean disc values projected into the local and distant
samples.

\begin{table}
\caption{Mean$\pm$r.m.s. values of the differences between the initial
  and fitted parameters over the 36 simulated local and 36 simulated
  distant giant LSBs. All the parameters were let free during the
  fitting process.\label{Tab:fake_galfit}}
\begin{tabular}{|c|c|c|}\hline
Parameter & Mean (initial-fitted) & $\sigma$ (initial-fitted)\\ \hline
  & SDSS & GOODS \\ \hline
\multicolumn{3}{c}{Bulge} \\ \hline
$\mu_0$ (mag/arcsec$^2$) & 0.14$\pm$0.02 & 0.33$\pm$0.02  \\ \hline
$r_d$ (kpc) & -0.13$\pm$0.02 & -0.11$\pm$0.03 \\ \hline
$n$ & -0.12$\pm$0.01& -0.18$\pm$0.02 \\ \hline
\multicolumn{3}{c}{Disc} \\ \hline
$\mu_0$ (mag/arcsec$^2$) & -0.07$\pm$0.04 & -0.06$\pm$0.04  \\ \hline
$r_d$ (kpc) & -0.58$\pm$0.15 & -0.77$\pm$0.25 \\ \hline
\end{tabular}
\end{table}

Tab.~\ref{Tab:fake_galfit} shows that the extended LSB discs can be
well recovered both in the local and distant samples. The only
significant effects are a slight underestimation of the bulge central
surface brightness (in particular for the distant sample), while it is
accurately recovered for the giant LSB discs, and a trend for the
giant LSB discs to be underestimated in size both in the local and
distant samples. We conclude that the morphological decomposition
described in Sect. 3.2 provide robust identifications of faint giant
LSB discs, and that the lack of detection of giant LSBs in the local
sample cannot be due to differences in surface brightness limits
between the two surveys, as expected from App. \ref{app:snr}.


\section{The SDSS $frac\_DeV$ parameter vs. the bulge-to-total light ratio.}

\label{app:fracdev}

\begin{figure}
	\includegraphics[width=80mm]{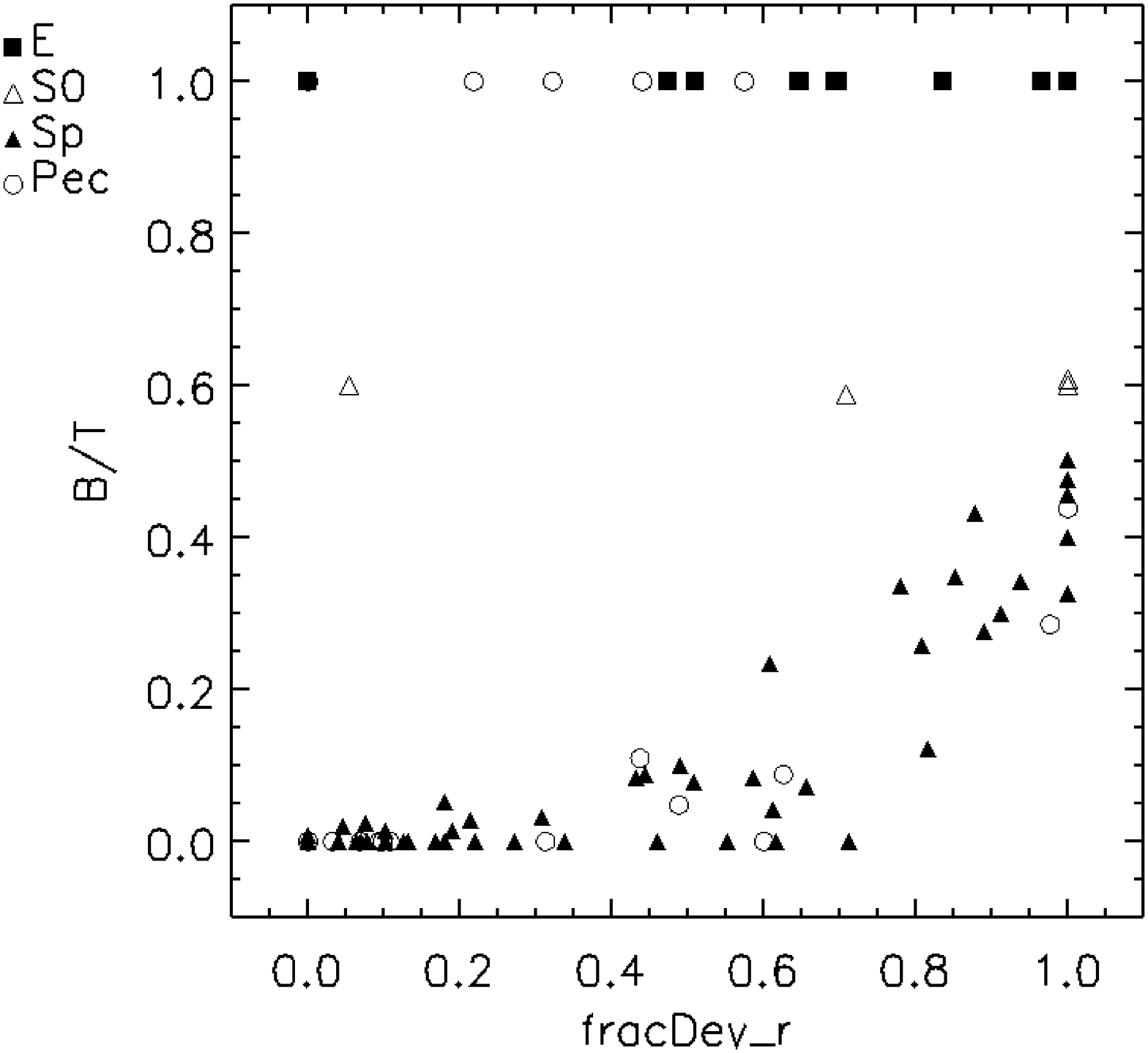}
	\caption{Comparison between the bulge-to-total light ratio and
          the SDSS $frac\_deV$ parameter for the local sub-sample of
          galaxies with $b/a>0.5$, in $r$ band.}
\label{compareSDSS}	
\end{figure}

The reduction pipeline of the SDSS database (\citealt{Stoughton02})
produces an automatic fit of each galaxy profile. Their code fits two
models in each band: a pure de Vaucouleurs profile truncated at
$7r_{eff}$, and a pure exponential profile truncated at $3r_{eff}$.
Their fit takes into account the PSF modelled by a double Gaussian.
The choice of fitting only one component instead of a more complicated
model is motivated by the huge amount of data and the computational
expense of such a process. A linear combination of the best-fit de
Vaucouleurs and exponential models is then re-fitted to the galaxy
image. The fraction of the de Vaucouleurs profile in the resulting fit
is recorded as the $frac\_deV$ parameter. In many papers this
parameter is used to distinguish between early and late-type galaxies,
or even as a proxy for the bulge-to-total light ratio (an elliptical
galaxy is expected to have $frac\_deV>0.5$, while a pure disc galaxy
is expected to have $frac\_deV < 0.5$, see, e.g.,
\citealt{Masters2010, Zhong12} and references therein). We compare in
Fig.~\ref{compareSDSS} the B/T ratio derived from our full bulge/disc
decomposition and the $frac\_deV$ parameter provided by the SDSS
database in $r$ band. We find no clear correlation between the two
parameters, although it can be noticed that the disc galaxies in the
sample with B/T$<$0.1 all have $frac\_deV<0.4$, and that except for
one object, all the elliptical galaxies have $frac\_deV>0.45$. The
spiral galaxies in the sample with 0.2$<$B/T$<$0.5 all have
$frac\_deV>0.8$, which means that they could be classified as S0 or E
galaxies using this criterion only.

\bibliographystyle{mn2e}
\bibliography{distlsb}

\label{lastpage}

\end{document}